\documentclass[aps,prl,twocolumn,superscriptaddress]{revtex4-2}
\usepackage{amsmath,amsfonts,amssymb}
\usepackage{graphicx,float,calc}
\usepackage{color,bm}
\usepackage{ulem}
\usepackage{braket}
\usepackage[colorlinks,urlcolor=blue,citecolor=blue,linkcolor=blue]{hyperref}

\begin{document}

\title{Arbitrarily Configurable Wavefunctions via Imaginary Gauge Phase Imprint in Non-Hermitian Lattices}

\author{Ji-Long Dong}
\affiliation{Key Laboratory of Atomic and Subatomic Structure and Quantum Control (Ministry of Education), Guangdong Basic Research Center of Excellence for Structure and Fundamental Interactions of Matter, South China Normal University, Guangzhou 510006, China}
\affiliation{Guangdong-Hong Kong Joint Laboratory of Quantum Matter, Frontier Research Institute for Physics, School of Physics, South China Normal University, Guangzhou 510006, China}

\author{Shi-Liang Zhu}
\affiliation{Key Laboratory of Atomic and Subatomic Structure and Quantum Control (Ministry of Education), Guangdong Basic Research Center of Excellence for Structure and Fundamental Interactions of Matter, South China Normal University, Guangzhou 510006, China}
\affiliation{Guangdong-Hong Kong Joint Laboratory of Quantum Matter, Frontier Research Institute for Physics, School of Physics, South China Normal University, Guangzhou 510006, China}
\affiliation{Quantum Science Center of Guangdong-Hong Kong-Macao Greater Bay Area (Guangdong), Shenzhen 518045, China}
\affiliation{Hefei National Laboratory, Hefei 230088, China}

\author{Dan-Wei Zhang}\email{danweizhang@m.scnu.edu.cn}
\affiliation{Key Laboratory of Atomic and Subatomic Structure and Quantum Control (Ministry of Education), Guangdong Basic Research Center of Excellence for Structure and Fundamental Interactions of Matter, South China Normal University, Guangzhou 510006, China}
\affiliation{Guangdong-Hong Kong Joint Laboratory of Quantum Matter, Frontier Research Institute for Physics, School of Physics, South China Normal University, Guangzhou 510006, China}
\affiliation{Quantum Science Center of Guangdong-Hong Kong-Macao Greater Bay Area (Guangdong), Shenzhen 518045, China}

\date{\today}

\begin{abstract}
We propose a general framework, termed the imaginary gauge phase imprint (IGPI), which enables engineering arbitrarily configurable wavefunctions with exact solutions and self-organization dynamics in any-dimensional non-Hermitian lattices under imaginary gauge fields. Using this method, we uncover a novel phase with exact critical wavefunctions, dubbed the skin critical phase (SCP), which is marked by unconventional localization, topological-skin, and dynamical characteristics.
Furthermore, we validate the IGPI by imprinting and visualizing complex fractal states with Sierpi\'{n}ski-carpet and Koch-snowflake profiles, as well as exotic super-moir\'{e} and 3D-moir\'{e} states in regular lattices. Our work not only offers fresh insights into non-Hermitian critical and fractal physics, but also provides a rigorous paradigm for controlling and visualizing wavefunction patterns using the IGPI in engineered non-Hermitian systems.
\end{abstract}

\maketitle

{\color{blue}\textit{Introduction.---}}Engineering complex wavefunctions of quantum particles or classical waves has long been a goal for exploring phases of matter and advancing quantum and wave-manipulation technologies. In localization and fractal physics \cite{Anderson1958,Jagannathan2021,Evers2008RMP,Gefen1980}, wavefunctions with fractal structures exhibit at extended-localized transition points or critical phases of quasiperiodic systems \cite{Aubry1980,Hofstadter1976,JHan1994,Chang1997,FLiu2015,JWang2016,QYao2024}. However, it is a challenge to rigorously confirm multifractal critical states in finite-size systems due to their general absence of exact wavefunctions and dynamical visualization \cite{HYao2019,YWang2020,YWang2021,Goncalves2024,XJLiu2024,YWang2022,SLi2026,XZhou2026}, even though the precise phase boundaries have been determined in several exactly solvable 1D quasiperiodic models \cite{XCZhou2023,Miguel2023,Migue2023b,Banerjee2025} and experimental efforts have been devoted to observing critical phases \cite{Richardella2010,Rispoli2019,goblot2020emergence,TXiao2021,HLi2023,Shimasaki2024,XLi2024,WHuang2025}. Recent advances in moir\'{e} lattices \cite{Nuckolls2025,PWang2019,ZMeng2023,XTWang2026} and non-Hermitian systems \cite{Bender1998,Ueda2020,Bergholtz2021} have in parallel enriched wavefunction configurations, including moir\'{e} quasicrystal patterns, delocalization and erratic localization induced by imaginary gauge potentials, \cite{Hatano1996,Hatano1997,Longhi2025,Nan2026} and the non-Hermitian skin effect (NHSE) \cite{SYao2018,Lee2016,Jin2019,Okuma2020,Borgnia2020,KZhang2020,LLi2020,lin2023topological,LXue2020,Helbig2020,Weidemann2020,XZhang2021,Liang2022,JWu2025,Han2026,JTao2026,HNie2026}. Within non-Hermitian quasiperiodic lattices, fractal spectrum can be extended to the complex energy plane \cite{LWang2024,JSun2024}, and 1D critical states exhibit boundary- or size-dependent localization \cite{lztang2021,XCai2022,SZLi2024,GJLiu2024,HLiang2025,Chakrabarty2025,YPWang2025}. Notably, the NHSE in low-dimensional topological lattices has been explored to control the wavefunction profiles of topological modes \cite{WWang2022,ZLin2026}. A fundamental question that remains unexplored concerns the non-Hermitian shaping of arbitrary wavefunctions for precisely creating exotic phases and manipulating wave patterns.

In this Letter, we address this question by developing a general framework for imprinting arbitrarily configurable wavefunctions in any-dimensional non-Hermitian lattices, termed the imaginary gauge phase imprint (IGPI), based on our exact solutions of the generalized Hatano-Nelson model with experimentally tunable imaginary gauge potentials \cite{Hatano1996,Hatano1997,LXue2020,Helbig2020,Weidemann2020,XZhang2021,Liang2022,JWu2025,Han2026,JTao2026,HNie2026}. The IGPI enables the self-organized formation of desired wavefunction profiles with ballistic wavepacket dynamics. Using this method, we uncover a novel phase with exact critical wavefunctions, namely the skin critical phase (SCP), featuring unique static, dynamical, and skin characteristics distinct from both conventional critical phase (CCP) \cite{JHan1994,Chang1997,FLiu2015,JWang2016,QYao2024,lztang2021,XCai2022,SZLi2024,GJLiu2024,HLiang2025} and NHSE \cite{SYao2018,Lee2016,Jin2019,Okuma2020,Borgnia2020,KZhang2020,LLi2020,lin2023topological}. 
Moreover, we validate the IGPI by imprinting and visualizing complex wavefunctions in higher dimensions, including fractal states with Sierpi\'{n}ski-carpet and Koch-snowflake profiles in non-fractal lattices, as well as super-moir\'{e} and 3D moir\'{e} states in untwisted lattices. Our work with exact solutions opens avenues for rigorously exploring exotic fractal and critical phases, and our proposed IGPI establishes a new paradigm to shape arbitrary wavefunction profiles, offering great capacity for quantum and classical wave manipulations in engineered non-Hermitian systems.

{\color{blue}\textit{Model and exact wavefunctions.---}}We consider a generalized non-Hermitian Hatano-Nelson model \cite{Hatano1996,Hatano1997} with spatially varying imaginary gauge phases \cite{Claes2021,ZQZhang2023,Midya2024,Longhi2025} in a $d$-dimensional regular lattice of size $N=L^d$ (with $L$ being the length along each direction), described by the tight-binding Hamiltonian
\begin{equation} \label{eq1}
\hat{H} = \sum_{\alpha=1}^{d}\sum_{\mathbf{r}} J\left[ e^{-g_{\mathbf{r}}^{(\alpha)}} \hat{c}_{\mathbf{r}}^{\dagger} \hat{c}_{\mathbf{r} + \mathbf{e}_\alpha} + e^{g_{\mathbf{r}}^{(\alpha)}} \hat{c}_{\mathbf{r} + \mathbf{e}_\alpha}^{\dagger} \hat{c}_{\mathbf{r}} \right] + \hat{H}_B.
\end{equation}
Here, $\hat{c}_{\mathbf{r}}^{\dagger}$ ($\hat{c}_{\mathbf{r}}$) denotes the creation (annihilation) operator at lattice site $\mathbf{r}$, $\mathbf{e}_\alpha$ is the unit vector along $\alpha$ direction, $J$ is the hopping strength, and $g_{\mathbf{r}}^{(\alpha)}$ represents imaginary gauge phases that generally depend on site $\mathbf{r}$ and nonreciprocal hopping direction $\alpha$. Hereafter, we set $J =\hbar= 1$ and $\hbar/J=1$ as the energy and time units, respectively.

In 1D case of $d=\alpha=1$, one can rewrite $g_{\mathbf{r}}^{(1)} \equiv g_n$ with the site $n=1,2,...,L$. For higher dimensions $d\geqslant2$, we first focus on the general case with coupled imaginary gauge field and consider the degenerated case later. To ensure exact solvability, we reformulate the imaginary gauge field $g_{\mathbf{r}}^{(\alpha)}$ as the difference between the cumulative imaginary gauge phases at neighboring sites along the $\alpha$ direction, i.e., $g_{\mathbf{r}}^{(\alpha)}\mapsto X_{\mathbf{r}+\mathbf{e}_\alpha} - X_{\mathbf{r}}$, where $X_{\mathbf{r}} = \sum_{l_1=1}^{r_1} \cdots \sum_{l_d=1}^{r_d} g_{\mathbf{l}}$ is the $d$-dimensional cumulative imaginary gauge phase. $\hat{H}_B$ controls the boundary condition: $\hat{H}_B = 0$ for the OBC, whereas for the PBC, $\hat{H}_B$ preserves nonreciprocal hopping between two boundary sites along each direction (see Supplemental Material (SM) Sec. A \cite{SM}). In this general case, the single-particle eigen-equation derived from the Hamiltonian (\ref{eq1}) is 
\begin{equation}\label{eq2}
       E \psi_{\mathbf{r}}  = \sum_{\alpha=1}^{d}  J \Big[e^{X_{\mathbf{r}} - X_{\mathbf{r}-\mathbf{e}_\alpha}} \psi_{\mathbf{r}-\mathbf{e}_\alpha} + e^{X_{\mathbf{r}}-X_{\mathbf{r} +\mathbf{e}_\alpha}} \psi_{\mathbf{r}+\mathbf{e}_\alpha} \Big],
\end{equation}
where $\psi_{\mathbf{r}}$ and $E$ are the eigenstate and eigenenergy.

By an imaginary gauge transformation \cite{Midya2024,Longhi2025}: $\psi_{\mathbf{r}} = \phi_{\mathbf{r}} \exp\left( X_{\mathbf{r}} \right)$, the gauge phase can be eliminated. 
This leads to a transformation of the non-Hermitian eigen-equation (\ref{eq2}) into its equivalent Hermitian form: $E \phi_{\mathbf{r}} = \sum_{\alpha=1}^{d} J \left( \phi_{\mathbf{r} - \mathbf{e}_\alpha} + \phi_{\mathbf{r} + \mathbf{e}_\alpha} \right)$, which allows exact solutions via separation of variables along all directions. However, under PBC, separation of variables requires the cumulative imaginary phase at the boundary to be identical for all directions, satisfying $X_{\{\mathbf{r} \mid r_\alpha = L\}}=X_{L}$. Letting $\phi_{\mathbf{r}} = \prod_{\alpha=1}^{d} \phi^{(\alpha)}_{r_\alpha}$ and $E = \sum_{\alpha=1}^{d} E^{(\alpha)}$, we obtain independent 1D equations along each direction $E^{(\alpha)} \phi^{(\alpha)}_{r_\alpha} = J ( \phi^{(\alpha)}_{r_\alpha - 1} + \phi^{(\alpha)}_{r_\alpha + 1} )$, where $\phi^{(\alpha)}_0 = \phi^{(\alpha)}_{L+1} = 0$ for the OBC and $\phi^{(\alpha)}_{r_\alpha+L} =\phi^{(\alpha)}_{r_\alpha} \exp(-X_L) = \phi^{(\alpha)}_{r_\alpha}\exp(-\bar{g}^{(\alpha)}L)$ for the PBC, with $\bar{g}^{(\alpha)}=\frac{1}{L}\sum_{l=1}^{L} g_l^{(\alpha)}$. 
\begin{figure}[h]
    \centering
    \includegraphics[width=0.95\linewidth]{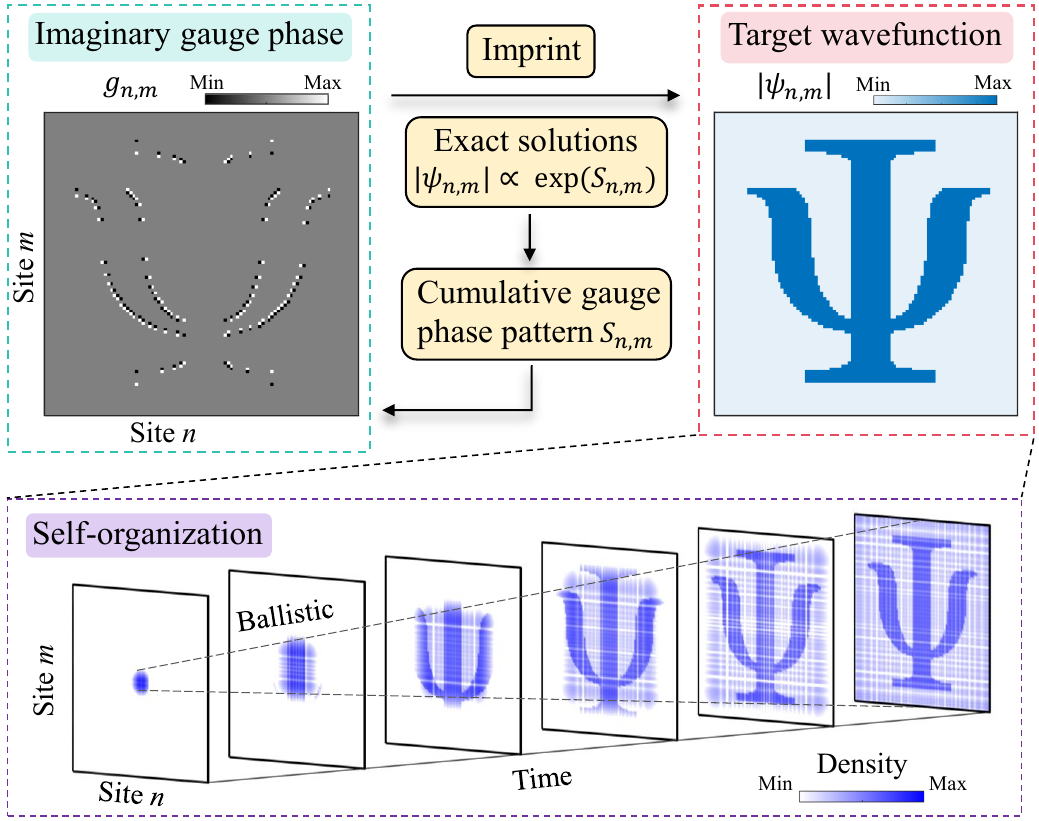}
    \caption{Schematic illustration of the IGPI framework. For a target wavefunction with an arbitrary pattern, e.g. $|\psi_{n,m}|$ with patten described by the letter ``$\Psi$'' in a 2D lattice, one can use exact solutions to obtain the cumulative gauge phase pattern $S_{n,m}$. The required imaginary gauge phase $g_{n,m}$ can be reconstructed, which imprints the desired wavefunction profile. The density patten ``$\Psi$'' can be self-organizational formed via free expansion began from a single-site excitation.}
    \label{fig1}
\end{figure}
By solving these equations and applying the inverse gauge transformation, we obtain exact wavefunctions (SM Sec. A \cite{SM}):  
\begin{equation}\label{eq3}
\psi_{\mathbf{r}}^{\{j^{(\alpha)}\}}=
\begin{cases}
      \prod_{\alpha=1}^{d} \sin[ \pi j^{(\alpha)} r_\alpha/(L+1) ] e^{X_{\mathbf{r}}}, & \text{OBC}; \\[2ex]
      \prod_{\alpha=1}^{d} e^{i 2\pi j^{(\alpha)} r_\alpha/L - \bar{g}^{(\alpha)} r_\alpha+X_{\mathbf{r}}}, & \text{PBC},
\end{cases}
\end{equation}
up to a normalization constant, where $\{j^{(\alpha)}\}$ denotes the eigenstate index with total number $N=L^d$ and $j^{(\alpha)} = 1, 2, \dots, L$. The corresponding eigenenergies are $E_{\text{OBC}}^{\{j^{(\alpha)}\}}=\sum_{\alpha=1}^{d} 2J \cos[\pi j^{(\alpha)}/(L+1)]$ and $E_{\text{PBC}}^{\{j^{(\alpha)}\}}=\sum_{\alpha=1}^{d}2J \cos[ 2\pi j^{(\alpha)}/L + i \bar{g}^{(\alpha)} ]$.

{\color{blue}\textit{IGPI framework.---}}A key observation on exact wavefunctions given by Eq. (\ref{eq3}) is the profile of $X_{\mathbf{r}}$, controlled by the imaginary gauge phase $g_{\mathbf{r}}^{(\alpha)}$. It provides a versatile method to imprint wavefunction profiles in this non-Hermitian lattice through designing $g_{\mathbf{r}}^{(\alpha)}$, termed the IGPI since it is similar to the \textit{real} phase imprint for wavefunction engineering of atomic gases \cite{Burger1999,Dobrek1999,Denschlag2000}. The IGPI method enables the creation of wavefunctions with arbitrary configurations in any dimension, which is schematically illustrated in Fig. \ref{fig1}. For a target wavefunction with a desired amplitude pattern, such as $|\psi_{n,m}|$ with pattern as the letter ``$\Psi$'' in a 2D lattice, one can use exact solutions of $\psi_{n,m}$ to obtain the corresponding cumulative gauge phase $X_{\mathbf{r}} \equiv S_{n,m}$. Then the required imaginary gauge phase can be constructed via the discrete difference relation $g_{\mathbf{r}}^{(\alpha)}\equiv g_{n,m}=S_{n,m}-S_{n-1,m}-S_{n,m-1}+S_{n-1,m-1}$ (see SM Sec. B \cite{SM}), with the distributions of $g_{n,m}$ and $|\psi_{n,m}|$ shown in the left and right sides of Fig.~\ref{fig1}, respectively.

\begin{figure}[htbp]
    \centering
    \includegraphics[width=0.95\linewidth]{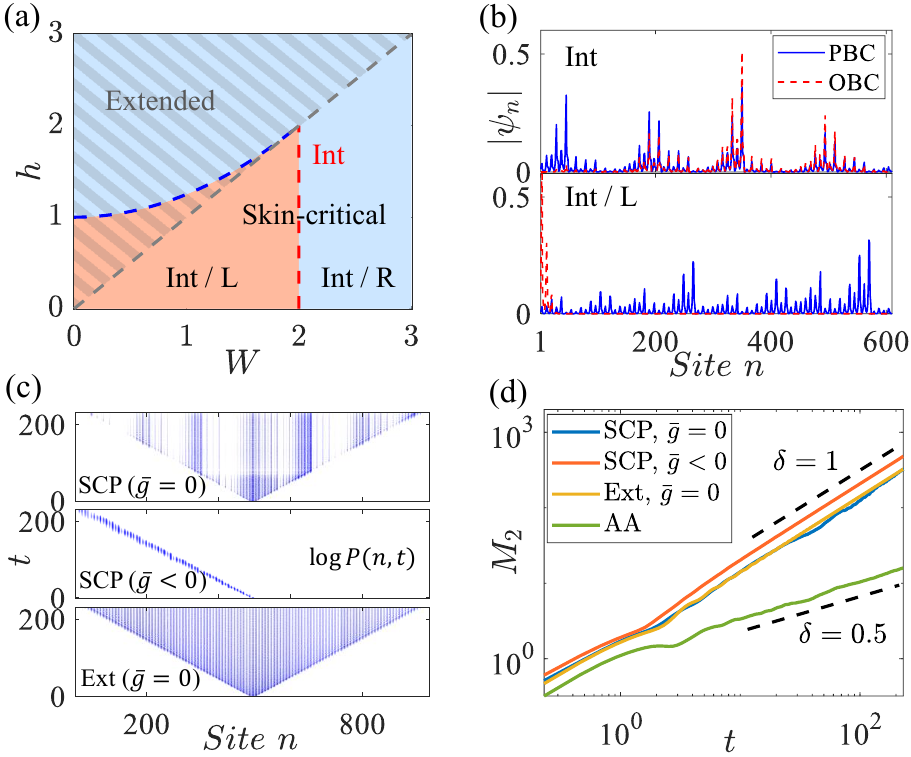}
    \caption{1D SCP. (a) Phase diagram on the $W$-$h$ plane. The system exhibits the SCP below the diagonal (grey dashed line) and extended phase in the striped region above. In the SCP, all eigenstates are skinned to bulk interfaces (Int) under the PBC for any winding number $\omega$; while under the OBC, they become interface, left (L) or right (R) boundary skin modes for $\omega=0,1,-1$, respectively. The orange and blue shading regions has $\omega=\pm1$ for $\bar{g}<0$ and $\bar{g}>0$, separating by exact boundaries as blue and red dashed lines with $\omega=\bar{g}=0$. (b) Wavefunction distributions $|\psi_n|$ under the PBC and OBC for $\omega=0$ ($W=2$ and $h=1$) and $\omega=1$ ($W=1.5$ and $h=1$). (c) Expansion dynamics with time-evolved density profiles $\log P(n,t)$ in the SCP ($\bar{g}=0$ and $\bar{g}<0$) and the extended (Ext) phase from top to bottom panels. (d) Corresponding time-evolved second moment $M_2(t)$ with ballistic scaling $\delta=1$. The simulated $M_2(t)$ at the critical point of the AA model with diffusive scaling $\delta=0.5$ is shown for comparison.}
    \label{fig2}
\end{figure}

\begin{figure}[t]
    \centering
    \includegraphics[width=0.95\linewidth]{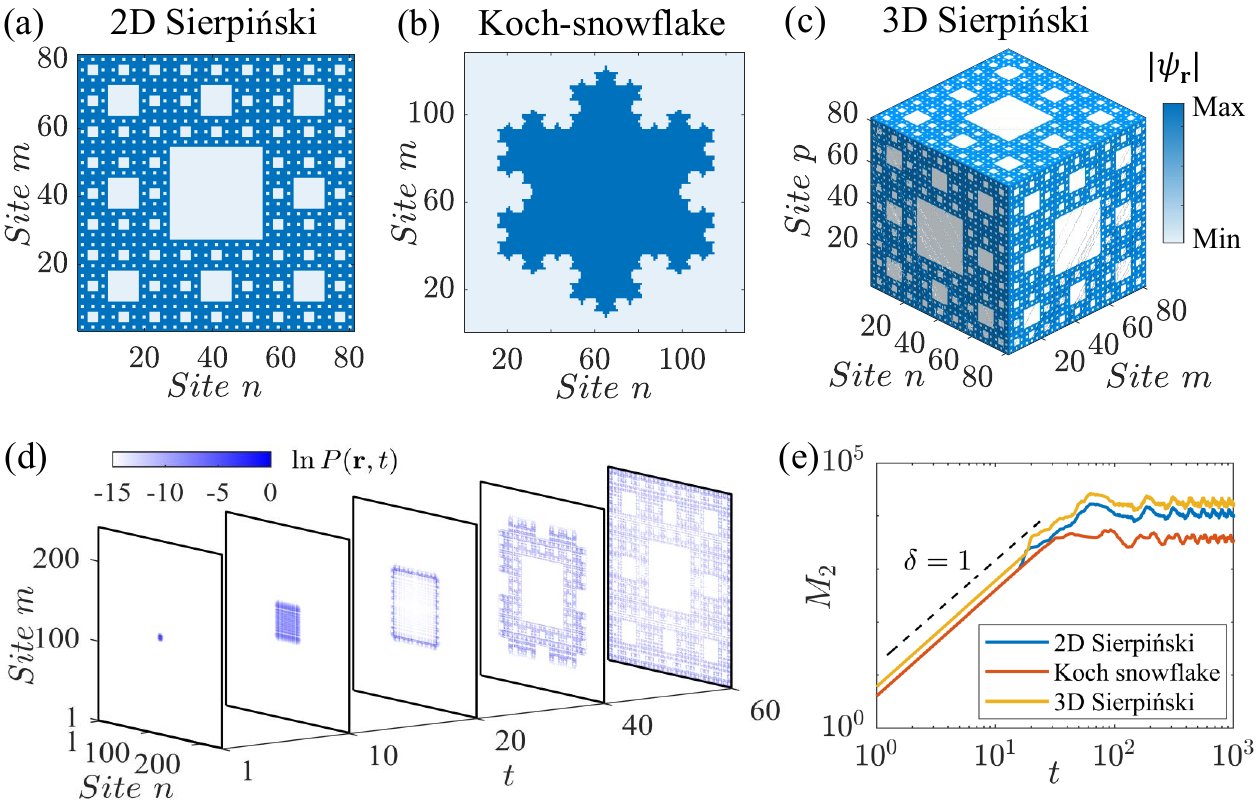}
    \caption{Imprinting and seeing fractals. (a) 2D Sierpi\'{n}ski-carpet, (b) Koch-snowflake, and (c) 3D Sierpi\'{n}ski-carpet wavefunctions in simple lattices under the PBC. (d) Seeing Sierpiński-carpet pattern from time slices of density distributions. (d) $M_2(t)$ during formation of fractal states in (a-c).}
    \label{fig3}
\end{figure}

We also study the expansion dynamics by considering a localized excitation $|\psi(0)\rangle=\hat{c}_{\mathbf{r}_0}^{\dagger}|0\rangle$ initially at the center $\mathbf{r}_0$ of an unoccupied lattice $|0\rangle$ at time $t=0$. The state at time $t$ is given by $|\psi(t)\rangle = e^{-i\hat{H}t}|\psi(0)\rangle /\|e^{-i\hat{H}t}|\psi(0)\rangle\|$, and the time-dependent single-particle density distribution is given by $P(\mathbf{r},t)=\langle\psi(t)|\hat{c}_{\mathbf{r}}^{\dagger}\hat{c}_{\mathbf{r}}|\psi(t)\rangle$, where the site $\mathbf{r}=n,(n,m),(n,m,p)$ for 1D, 2D, 3D regular lattices. By simulating the expansion dynamics, we find the self-organizational formation of the pattern of the target wavefunction with ballistic dynamics (see Figs. \ref{fig2}(d) and \ref{fig3}(e) for the ballistic dynamics in other cases), such as the example ``$\Psi$'' shown in Fig. \ref{fig1}. The self-organization behaviour here is due to the fact that all eigenstates share similar patterns from the cumulative imaginary phase $X_{\mathbf{r}}$ (see Eq. (\ref{eq3})). Thus, the IGPI establishes a general and dynamically fast framework to generate arbitrarily configurable wavefunctions in non-Hermitian lattices.

{\color{blue}\textit{SCP in 1D.---}}Using the IGPI method, we reveal a novel non-Hermitian critical phase. For clarity, we here focus on the 1D case (and discuss other cases latter), under the imprint phase $X_{\mathbf{r}} \doteq X_n=\sum_{l=1}^{n-1} g_l$ with the imaginary gauge phase $g_n = \ln(|W \cos(2 \pi \beta n + \varphi)+h \cos(\pi n)|)$. Here $\beta = (\sqrt{5}-1)/2$ is an irrational number, $W$ and $h$ are the strengths of the quasiperiodic and periodic modulations, respectively, and $\varphi$ is a sampling phase. The exact wavefunction of the $j$-th ($j^{(\alpha)}\doteq j$) eigenstate $\psi_n^{(j)}$ is give by Eq. (\ref{eq3}) via reducing $\bar{g}^{(\alpha)}\mapsto\bar{g}=\frac{1}{L}\sum_{n=1}^{L} g_n$. The corresponding eigenenergies is real for the OBC; while for the PBC, the spectrum can form an elliptical loop in the complex energy place, which is topologically characterized by the spectral winding number $\omega_j$ \cite{Gong2018} (see End Matter). When $\bar{g}<0$ and $\bar{g}>0$, one has $\omega_j=\omega=\pm1$ for any $j$-th eigenstate, respectively. For the globally reciprocal case  
$\bar{g}=0$, the PBC spectrum becomes real without loop structure akin to Hermitian cases and thus $\omega=0$. According to the correspondence between the winding number and NHSE \cite{Okuma2020,Borgnia2020,KZhang2020}, all OBC eigenstates are localized at the left and right edges of the 1D non-Hermitian system when $\omega=\pm1$, respectively, while there is no boundary skin modes for $\omega=0$. The phase diagram of the winding number in the $W$-$h$ plane is shown in Fig. \ref{fig2}(a), where two distinct regions with $\omega=\pm 1$ are separated by two exact topological boundaries: $W=W_c=2\sqrt{h-1}$ for $|W|<|h|$ and $W_c=2$ for $|W|\geqslant|h|$. At topological transitions, we extract the correlation-length and dynamical critical exponents as $\nu=z=1$ from the scaling analysis in SM Sec. C \cite{SM}.

We use the fractal dimension $D_f$ to quantify localization \cite{Evers2008RMP}. By computing $D_f$ under the PBC (see End Matter), we obtain the localization phase diagram in Fig. \ref{fig2}(a). For $W<h$, the system is in the extended phase with $D_f \approx 1$, whereas in the critical phase with $0<D_f<1$ for $W>h$, separating by the phase boundary $W=h$. In the critical phase, the average density distribution $\overline{\rho_n} = \frac{1}{L}\sum_{j=1}^L |\psi_n^{(j)}|^2\sim |\psi_n^{(j)}|^2$, such that all eigenstates exhibit macroscopically multifractal profiles in the system. While in the CCP, it is almost spatially uniform with $\overline{\rho_n} \sim 1/L$ as the eigenstates exhibit distinct multifractal distributions \cite{JHan1994,Chang1997,FLiu2015,JWang2016,QYao2024}. Under the OBC, the critical eigenstates become skin modes localized at the left (right) boundary for $\omega=1$ ($\omega=-1$) when $\bar{g}<0$ ($\bar{g}>0$), as shown in Fig. \ref{fig2}(b). Interestingly, in the globally reciprocal case of $\bar{g}=0$ with $\omega=0$, the critical eigenstates exhibit some main peaks in the lattice bulk for both OBC and PBC, which are located at the interfaces of $X_n$. Such interface-skin critical states also exhibit in non-reciprocal case of $\bar{g}\neq0$ under the PBC, whose peaks are around the interfaces of $X_n-\bar{g}n$. Due to these novel skin and localization properties, we term the critical phase as the SCP. The physical origin of the SCP is rooted in the extreme value statistics of random walks \cite{Schehr2012,Majumdar2020,Longhi2025}, under the quasiperiodic imaginary gauge field $g_n$. See End Matter for more details of the SCP.

We also study the expansion dynamics by considering a localized excitation initially at the center of the 1D lattice at time $t=0$. As shown in Fig. \ref{fig2}(c), in the SCP and extended phase with $\bar{g} =\omega=0$, the initial excitation exhibits an undirectional propagation and extends across the lattice with multifractal and periodic profiles, respectively, whereas it directionally propagates to the left for $\bar{g} < 0$ ($\omega=1$) as the skin dynamics. We characterize the wave expansion by using the second moment $M_2(t)= [\sum_{n=1}^{L} (n - L/2)^2 P(n,t)]^{1/2}$, which describes the wave-packet width. In the long-time evolution, one has $M_2(t) \sim t^\delta$, where the diffusion exponent $\delta$ takes $\delta=1$, $\delta=0$, and $\delta\approx 0.5$ corresponding to the ballistic, localization, and diffusive transports for conventional extended, localized, and critical phases, respectively \cite{YWang2020}. Fig. \ref{fig2}(d) shows corresponding $M_2(t)$ (averaged over sampling phase $\varphi$) in the SCP and extended phase, both of which exhibit the same wavepacket width during expansion and ballistic diffusion with $\delta\approx1$, in sharp contrast to the common wisdom that the critical and extended phases have different diffusion behaviours. The ballistic characteristics also distinguishes the SCP from the CCP with diffusive dynamics, such as that in the criticality of the Aubry-Andr\'{e} (AA) model \cite{Aubry1980}.

{\color{blue}\textit{Imprinting fractal and moir\'{e} wavefunctions.---}}We proceed to validate the IGPI method by engineering exotic wavefunctions in higher dimensions. We first create fractal states in 2D and 3D cases, where the imprint phase can be simplifies as $X_{\mathbf{r}} \equiv S_{n,m}=\sum_{l_1=1}^{n} \sum_{l_2=1}^{m} g_{l_1,l_2}$ and $X_{\mathbf{r}} \equiv V_{n,m,p} = \sum_{l_1=1}^{n} \sum_{l_2=1}^{m} \sum_{l_3=1}^{p} g_{l_1,l_2,l_3}$, respectively, with the site indices $\{n,m,p\}$ along three different directions. Under the IGPI procedure with proper $g_{n,m}$, we can create fractal wavefunctions with Sierpi\'{n}ski-carpet and Koch-snowflake profiles in 2D square lattices, respectively, as shown in Figs. \ref{fig3}(a) and \ref{fig3}(b), respectively. They are basic examples of geometric fractals with self-similarity and non-integer fractal dimension, such as $D_f\approx1.93$ for the 2D Sierpi\'{n}ski wavefunction and $D_f\approx1.29$ for the outline of the snowflake wavefunction. The Sierpi\'{n}ski states can be extended to 3D cubic lattices with proper $g_{n,m,p}$, as shown that in Fig. \ref{fig3}(c) and others in SM Sec. B \cite{SM}. Moreover, the patterns of fractal wavefunctions can be visualized from the real-time wave-packet expansion, as shown in Fig. \ref{fig3}(d). Furthermore, we find that the time-evolved second moment $M_2(t)$ for all three cases grows linearly with time $t$ before saturation for finite-size lattices,
as shown in Fig.~\ref{fig3}(e). This indicates the ballistic dynamics in the formation of fractal wavefunctions in regular lattices, in contrast to the sub-ballistic transport in geometrically fractal lattices \cite{XYXu2021,Biesenthal2022,RojoFrancas2024}. This ballistic behavior allow us to fast and dynamically visualize fractal pattern of wavefunctions in engineered non-Hermitian systems \cite{LXue2020,Helbig2020,Weidemann2020,XZhang2021,Liang2022,Han2026,JTao2026,HNie2026,JXZhong2026,Longhi2025}.

\begin{figure}[t]
    \centering
    \includegraphics[width=0.85\linewidth]{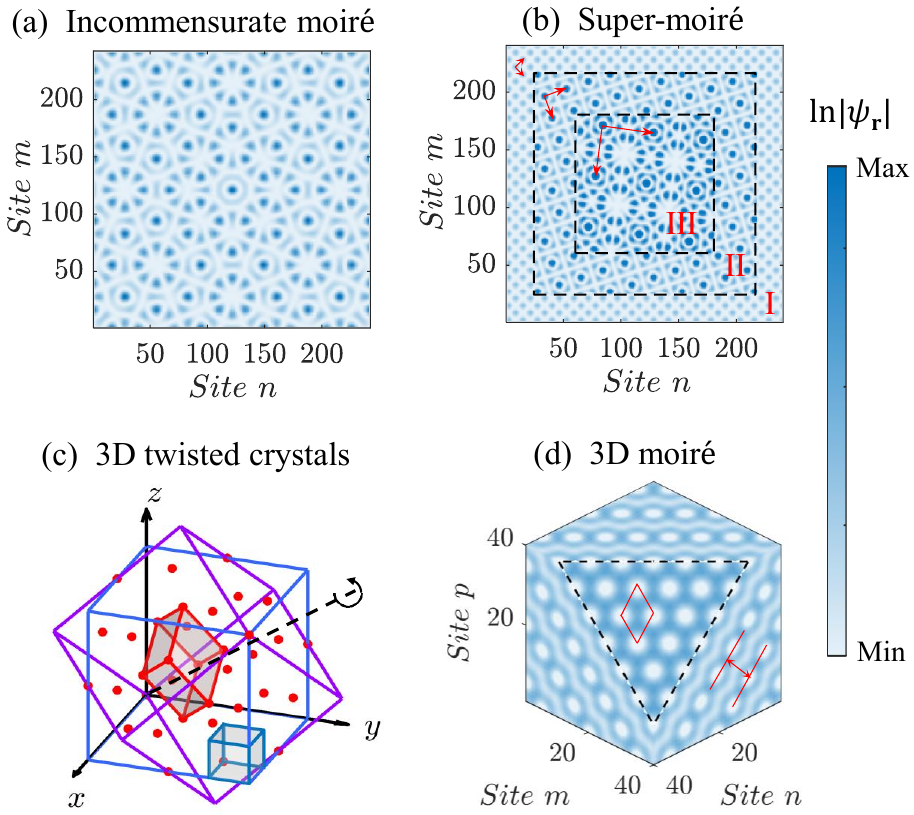}
    \caption{Imprinting moir\'{e} states under the PBC. (a) 2D incommensurate moir\'{e} and (b) super-moir\'{e} wavefunctions in square lattices. Regions I, II and III in (b) show square, moir\'{e}, and super-moir\'{e} patterns (with relative rotations denoted by red arrows), respective. (c) Illustration of 3D twisted crystals. (d) 3D moir\'{e} wavefunction in a cubic lattice.}
    \label{fig4}
\end{figure}

We then use the IGPI method to engineer moir\'{e} states \cite{Nuckolls2025,PWang2019,ZMeng2023,Ren2025,Wang2019,Wang2024,Wang2025}, with desired moir\'{e} patterns imprinted into wavefunctions within regular lattices without moir\'{e} potentials (see SM Sec. B for moir\'{e} lattice potentials \cite{SM}). For instance, we shape the wavefunction with an incommensurate moir\'{e} profile in a 2D square lattice, as shown in Fig.~\ref{fig4}(a), which mimics the moir\'{e} quasicrystal formed by superimposing two layers of square potentials with a twisted angle $\theta = \pi/4$ \cite{PWang2019}. When the twist angle of two layers is a Pythagorean angle (e.g., $\theta = \arctan (3/4)$), a moir\'{e} periodic potential an be formed. If a third layer potential is added and the rotation angles between adjacent layers keep commensurate, the three-layer superposition further interferes to form a super-moir\'{e} potential \cite{Ren2025,Wang2019}. Such a super-moir\'{e} pattern can also imprinted into the wavefunction profile in a square lattice, as demonstrated in Fig. \ref{fig4}(b). Furthermore, the non-Abelian rotational characteristics of 3D lattices can give rise to richer moir\'{e} physics \cite{Wang2024,Wang2025}. When two 3D cubic lattices undergo a relative rotation along the body diagonal under the commensurate condition (see Fig.~\ref{fig4}(c)), a subset of lattice sites (red dots) coincide to form a 3D moir\'{e} crystal, and the original small-period cubic unit cell (blue wireframe) transforms into a larger trigonal unit cell (red wireframe) \cite{Wang2024,Wang2025}. Fig.~\ref{fig4}(d) illustrates the imprinted wavefunction with such a complex 3D moir\'{e} pattern in a cubic lattice, where the cross-sectional and surface profiles mimic the geometric morphology of the trigonal unit cell.

{\color{blue}\textit{Discussion and conclusion.---}}We note that the SCP revealed in 1D case can be directly extended to higher dimensions and other non-Hermitian systems. In our SM Sec. D \cite{SM}, we reveal the 2D SCP with exact wavefunctions and bulk-interface, line-boundary, and corner skin modes. Adding an on-site quasiperiodic potential can induce the localization transition between the SCP and localized phase, and can enlarge the regime of the SCP with interface-skin modes (SM Sec. E \cite{SM}). Moreover, similar skin-critical states can exhibit in open quantum systems described by a Lindblad master equation (SM Sec. F \cite{SM}), where the effective nonreciprocal hopping arises from the dissipative system–environment couplings.

In summary, we have proposed the IGPI as a general and rigorous mechanism to engineer arbitrarily configurable wavefunctions with self-organization dynamics in any-dimensional non-Hermitian lattices. We have used the IGPI to uncover the SCP with properties different from those in conventional NHSE and critical phases. We have further validated the IGPI by imprinting and visualizing wavefunctions with complex fractal and moir\'{e} profiles. 
Our work opens new avenues for exploring novel phases of matter and manipulating quantum and classical waves with engineered non-Hermiticity. It also invites further extension of the IGPI method to many-body interaction regimes and non-Abelian imaginary gauge fields for investigating more exotic localization and topological phases \cite{YWang2020,YWang2021,Goncalves2024,DWZhang2020,Shimomura2024,Gliozzi2024,Yoshida2024,YQin2025,Chakrabarty2025b,ZPang2024,XChen2026,RJChen2025,Sanahal2025,WZhang2025,YMiao2025,DW2020TAI}. Since uniform and disordered imaginary gauge potentials have been realized to observe the NHSE \cite{LXue2020,Helbig2020,Weidemann2020,XZhang2021,Liang2022,JWu2025,Han2026,JTao2026,HNie2026} and the erratic skin localization \cite{JXZhong2026,Longhi2025} in a range of experimental platforms, we expect that our IGPI method will soon be achieved therein.

\textit{Note added.---}Recently, we noticed a preprint on the quasiperiodic skin criticality in a 1D non-Hermitian lattice \cite{ZChen2026}, similar as the 1D SCP. In our present work, we reveals the phase diagram, ballistic dynamics and higher-dimensional extensions of the SCP. Furthermore, we establish a general mechanism to imprint arbitrary wavefunctions in any-dimensional non-Hermitian lattices.

\begin{acknowledgments}
The authors acknowledge stimulating discussions with Z. D. Wang. This work is supported by the National Key Research and Development Program of China (Grant No.  2022YFA1405300), the Innovation Program for Quantum Science and Technology (Grant No. 2021ZD0301705), the Guangdong Basic and Applied Basic Research Foundation (Grant No. 2024B1515020018), the Guangdong Provincial Quantum Science Strategic Initiative (Grants No. GDZX2304002 and No. GDZX2404001), and the Science and Technology Program of Guangzhou (Grant No. 2024A04J3004). 

\end{acknowledgments}

\bibliography{reference}

@article{Anderson1958,
	title = {Absence of Diffusion in Certain Random Lattices},
	author = {Anderson, P. W.},
	journal = {Phys. Rev.},
	volume = {109},
	issue = {5},
	pages = {1492--1505},
	numpages = {0},
	year = {1958},
	month = {Mar},
	publisher = {American Physical Society},
	doi = {10.1103/PhysRev.109.1492},
	url = {https://link.aps.org/doi/10.1103/PhysRev.109.1492}
}

@article{Aubry1980,
	title={Analyticity breaking and Anderson localization in incommensurate lattices},
	author={Aubry, Serge and Andr{\'e}, Gilles},
	journal={Ann. Israel Phys. Soc},
	volume={3},
	number={133},
	pages={18},
	year={1980}
}

@article{goblot2020emergence,
	title={Emergence of criticality through a cascade of delocalization transitions in quasiperiodic chains},
	author={Goblot, Valentin and {\v{S}}trkalj, Antonio and Pernet, Nicolas and Lado, Jose L and Dorow, C and Lema{\^\i}tre, Aristide and Le Gratiet, Luc and Harouri, Abdelmounaim and Sagnes, Isabelle and Ravets, Sylvain and others},
	journal={Nat. Phys.},
	volume={16},
	number={8},
	pages={832--836},
	doi={10.1038/s41567-020-0908-7},
	year={2020},
	publisher={Nature Publishing Group UK London}
}

@Article{DWZhang2020,
  author    = {Zhang, Dan-Wei and Chen, Yu-Lian and Zhang, Guo-Qing and Lang, Li-Jun and Li, Zhi and Zhu, Shi-Liang},
  journal   = {Phys. Rev. B},
  title     = {Skin superfluid, topological Mott insulators, and asymmetric dynamics in an interacting non-Hermitian Aubry-Andr\'e-Harper model},
  year      = {2020},
  month     = {Jun},
  pages     = {235150},
  volume    = {101},
  doi       = {10.1103/PhysRevB.101.235150},
  issue     = {23},
  numpages  = {9},
  publisher = {American Physical Society},
  url       = {https://link.aps.org/doi/10.1103/PhysRevB.101.235150},
}

@article{lztang2021,
	title = {Localization and topological transitions in non-Hermitian quasiperiodic lattices},
	author = {Tang, Ling-Zhi and Zhang, Guo-Qing and Zhang, Ling-Feng and Zhang, Dan-Wei},
	journal = {Phys. Rev. A},
	volume = {103},
	issue = {3},
	pages = {033325},
	numpages = {9},
	year = {2021},
	month = {Mar},
	publisher = {American Physical Society},
	doi = {10.1103/PhysRevA.103.033325},
	url = {https://link.aps.org/doi/10.1103/PhysRevA.103.033325}
}

@Article{XLi2024,
  author    = {Li, Xuegang and Xu, Huikai and Wang, Junhua and Tang, Ling-Zhi and Zhang, Dan-Wei and Yang, Chuhong and Su, Tang and Wang, Chenlu and Mi, Zhenyu and Sun, Weijie and Liang, Xuehui and Chen, Mo and Li, Chengyao and Zhang, Yingshan and Linghu, Kehuan and Han, Jiaxiu and Liu, Weiyang and Feng, Yulong and Liu, Pei and Xue, Guangming and Zhang, Jingning and Jin, Yirong and Zhu, Shi-Liang and Yu, Haifeng and Zhao, S. P. and Xue, Qi-Kun},
  journal   = {Phys. Rev. Res.},
  title     = {Mapping the topology-localization phase diagram with quasiperiodic disorder using a programmable superconducting simulator},
  year      = {2024},
  month     = {Nov},
  pages     = {L042038},
  volume    = {6},
  doi       = {10.1103/PhysRevResearch.6.L042038},
  issue     = {4},
  numpages  = {6},
  publisher = {American Physical Society},
  url       = {https://link.aps.org/doi/10.1103/PhysRevResearch.6.L042038},
}

@Article{YWang2021,
  author    = {Wang, Yucheng and Cheng, Chen and Liu, Xiong-Jun and Yu, Dapeng},
  journal   = {Phys. Rev. Lett.},
  title     = {Many-Body Critical Phase: Extended and Nonthermal},
  year      = {2021},
  month     = {Feb},
  pages     = {080602},
  volume    = {126},
  doi       = {10.1103/PhysRevLett.126.080602},
  issue     = {8},
  numpages  = {6},
  publisher = {American Physical Society},
  url       = {https://link.aps.org/doi/10.1103/PhysRevLett.126.080602},
}

@Article{Miguel2023,
  author    = {Gon\ifmmode \mbox{\c{c}}\else \c{c}\fi{}alves, Miguel and Amorim, Bruno and Castro, Eduardo V. and Ribeiro, Pedro},
  journal   = {Phys. Rev. Lett.},
  title     = {Critical Phase Dualities in 1D Exactly Solvable Quasiperiodic Models},
  year      = {2023},
  month     = {Nov},
  pages     = {186303},
  volume    = {131},
  doi       = {10.1103/PhysRevLett.131.186303},
  issue     = {18},
  numpages  = {6},
  publisher = {American Physical Society},
  url       = {https://link.aps.org/doi/10.1103/PhysRevLett.131.186303},
}

@Article{XCZhou2023,
  author    = {Zhou, Xin-Chi and Wang, Yongjian and Poon, Ting-Fung Jeffrey and Zhou, Qi and Liu, Xiong-Jun},
  journal   = {Phys. Rev. Lett.},
  title     = {Exact New Mobility Edges between Critical and Localized States},
  year      = {2023},
  month     = {Oct},
  pages     = {176401},
  volume    = {131},
  doi       = {10.1103/PhysRevLett.131.176401},
  issue     = {17},
  numpages  = {8},
  publisher = {American Physical Society},
  url       = {https://link.aps.org/doi/10.1103/PhysRevLett.131.176401},
}

@article{Hatano1996,
	title = {Localization Transitions in Non-Hermitian Quantum Mechanics},
	author = {Hatano, Naomichi and Nelson, David R.},
	journal = {Phys. Rev. Lett.},
	volume = {77},
	issue = {3},
	pages = {570--573},
	numpages = {0},
	year = {1996},
	month = {Jul},
	publisher = {American Physical Society},
	doi = {10.1103/PhysRevLett.77.570},
	url = {https://link.aps.org/doi/10.1103/PhysRevLett.77.570}
}

@article{Hatano1997,
	title = {Vortex pinning and non-Hermitian quantum mechanics},
	author = {Hatano, Naomichi and Nelson, David R.},
	journal = {Phys. Rev. B},
	volume = {56},
	issue = {14},
	pages = {8651--8673},
	numpages = {0},
	year = {1997},
	month = {Oct},
	publisher = {American Physical Society},
	doi = {10.1103/PhysRevB.56.8651},
	url = {https://link.aps.org/doi/10.1103/PhysRevB.56.8651}
}

@article{Bender1998,
	title = {Real Spectra in Non-Hermitian Hamiltonians Having PT Symmetry},
	author = {Bender, Carl M. and Boettcher, Stefan},
	journal = {Phys. Rev. Lett.},
	volume = {80},
	issue = {24},
	pages = {5243--5246},
	numpages = {0},
	year = {1998},
	month = {Jun},
	publisher = {American Physical Society},
	doi = {10.1103/PhysRevLett.80.5243},
	url = {https://link.aps.org/doi/10.1103/PhysRevLett.80.5243}
}

@article{Ueda2020,
	author = {Yuto Ashida, Zongping Gong and Masahito Ueda},
	title = {Non-Hermitian physics},
	journal = {Adv. Phys.},
	volume = {69},
	number = {3},
	pages = {249--435},
	year = {2020},
	publisher = {Taylor \& Francis},
	doi = {10.1080/00018732.2021.1876991}
}

@article{Bergholtz2021,
	title = {Exceptional topology of non-Hermitian systems},
	author = {Bergholtz, Emil J. and Budich, Jan Carl and Kunst, Flore K.},
	journal = {Rev. Mod. Phys.},
	volume = {93},
	issue = {1},
	pages = {015005},
	numpages = {31},
	year = {2021},
	month = {Feb},
	publisher = {American Physical Society},
	doi = {10.1103/RevModPhys.93.015005},
	url = {https://link.aps.org/doi/10.1103/RevModPhys.93.015005}
}

@article{Gong2018,
	title = {Topological Phases of Non-Hermitian Systems},
	author = {Gong, Zongping and Ashida, Yuto and Kawabata, Kohei and Takasan, Kazuaki and Higashikawa, Sho and Ueda, Masahito},
	journal = {Phys. Rev. X},
	volume = {8},
	issue = {3},
	pages = {031079},
	numpages = {33},
	year = {2018},
	month = {Sep},
	publisher = {American Physical Society},
	doi = {10.1103/PhysRevX.8.031079},
	url = {https://link.aps.org/doi/10.1103/PhysRevX.8.031079}
}

@article{Jin2019,
	title = {Bulk-boundary correspondence in a non-Hermitian system in one dimension with chiral inversion symmetry},
	author = {Jin, L. and Song, Z.},
	journal = {Phys. Rev. B},
	volume = {99},
	issue = {8},
	pages = {081103},
	numpages = {9},
	year = {2019},
	month = {Feb},
	publisher = {American Physical Society},
	doi = {10.1103/PhysRevB.99.081103},
	url = {https://link.aps.org/doi/10.1103/PhysRevB.99.081103}
}

@Article{LXue2020,
  author  = {Xiao, Lei and Deng, Tianshu and Wang, Kunkun and Zhu, Gaoyan and Wang, Zhong and Yi, Wei and Xue, Peng},
  journal = {Nat. Phys.},
  title   = {Non-Hermitian bulk–boundary correspondence in quantum dynamics},
  year    = {2020},
  issn    = {1745-2481},
  number  = {7},
  pages   = {761-766},
  volume  = {16},
  doi     = {10.1038/s41567-020-0836-6},
  url     = {https://doi.org/10.1038/s41567-020-0836-6},
}

@Article{SYao2018,
  author    = {Yao, Shunyu and Wang, Zhong},
  journal   = {Phys. Rev. Lett.},
  title     = {Edge States and Topological Invariants of Non-Hermitian Systems},
  year      = {2018},
  month     = {Aug},
  pages     = {086803},
  volume    = {121},
  doi       = {10.1103/PhysRevLett.121.086803},
  issue     = {8},
  numpages  = {8},
  publisher = {American Physical Society},
  url       = {https://link.aps.org/doi/10.1103/PhysRevLett.121.086803},
}

@Article{LLi2020,
  author    = {Li, Linhu and Lee, Ching Hua and Mu, Sen and Gong, Jiangbin},
  journal   = {Nat. Commun.},
  title     = {Critical non-Hermitian skin effect},
  year      = {2020},
  issn      = {2041-1723},
  number    = {1},
  pages     = {5491},
  volume    = {11},
  doi       = {10.1038/s41467-020-18917-4},
  publisher = {Nature Publishing Group UK London},
}

@Article{XZhang2021,
  author    = {Zhang, Xiujuan and Tian, Yuan and Jiang, Jian-Hua and Lu, Ming-Hui and Chen, Yan-Feng},
  journal   = {Nat. Commun.},
  title     = {Observation of higher-order non-Hermitian skin effect},
  year      = {2021},
  number    = {1},
  pages     = {5377},
  volume    = {12},
  doi       = {https://doi.org/10.1038/s41467-021-25716-y},
  publisher = {Nature Publishing Group UK London},
}

@article{Liang2022,
	title = {Dynamic Signatures of Non-Hermitian Skin Effect and Topology in Ultracold Atoms},
	author = {Liang, Qian and Xie, Dizhou and Dong, Zhaoli and Li, Haowei and Li, Hang and Gadway, Bryce and Yi, Wei and Yan, Bo},
	journal = {Phys. Rev. Lett.},
	volume = {129},
	issue = {7},
	pages = {070401},
	numpages = {6},
	year = {2022},
	month = {Aug},
	publisher = {American Physical Society},
	doi = {10.1103/PhysRevLett.129.070401},
	url = {https://link.aps.org/doi/10.1103/PhysRevLett.129.070401}
}

@article{lin2023topological,
	title={Topological non-Hermitian skin effect},
	author={Lin, Rijia and Tai, Tommy and Li, Linhu and Lee, Ching Hua},
	journal = {Front. Phys.},
	volume = {18},
	number = {5},
	pages = {53605},
	ISSN = {2095-0470},
	DOI = {10.1007/s11467-023-1309-z},
	url = {https://doi.org/10.1007/s11467-023-1309-z},
	year = {2023},
}

@article{DW2020TAI,
	title={Non-Hermitian topological Anderson insulators},
	author={Zhang, Dan-Wei and Tang, Ling-Zhi and Lang, Li-Jun and Yan, Hui and Zhu, Shi-Liang},
	journal = {Sci. China-Phys. Mech. Astron.},
	volume = {63},
	pages = {267062},
	DOI = {10.1007/s11433-020-1521-9},
	year = {2020},
}

@Article{LWang2024,
  author    = {Wang, Li and Wang, Zhenbo and Chen, Shu},
  journal   = {Phys. Rev. B},
  title     = {Non-Hermitian butterfly spectra in a family of quasiperiodic lattices},
  year      = {2024},
  month     = {Aug},
  pages     = {L060201},
  volume    = {110},
  doi       = {10.1103/PhysRevB.110.L060201},
  issue     = {6},
  numpages  = {8},
  publisher = {American Physical Society},
  url       = {https://link.aps.org/doi/10.1103/PhysRevB.110.L060201},
}

@Article{GJLiu2024,
  author    = {Liu, Gui-Juan and Zhang, Jia-Ming and Li, Shan-Zhong and Li, Zhi},
  journal   = {Phys. Rev. A},
  title     = {Emergent strength-dependent scale-free mobility edge in a nonreciprocal long-range Aubry-Andr\'e-Harper model},
  year      = {2024},
  month     = {Jul},
  pages     = {012222},
  volume    = {110},
  doi       = {10.1103/PhysRevA.110.012222},
  issue     = {1},
  numpages  = {9},
  publisher = {American Physical Society},
  url       = {https://link.aps.org/doi/10.1103/PhysRevA.110.012222},
}

@Article{SZLi2024,
  author    = {Li, Shan-Zhong and Li, Zhi},
  journal   = {Phys. Rev. B},
  title     = {Ring structure in the complex plane: A fingerprint of a non-Hermitian mobility edge},
  year      = {2024},
  month     = {Jul},
  pages     = {L041102},
  volume    = {110},
  doi       = {10.1103/PhysRevB.110.L041102},
  issue     = {4},
  numpages  = {8},
  publisher = {American Physical Society},
  url       = {https://link.aps.org/doi/10.1103/PhysRevB.110.L041102},
}

@Article{Midya2024,
  author    = {Midya, Bikashkali},
  journal   = {Phys. Rev. A},
  title     = {Topological phase transition in fluctuating imaginary gauge fields},
  year      = {2024},
  month     = {Jun},
  pages     = {L061502},
  volume    = {109},
  doi       = {10.1103/PhysRevA.109.L061502},
  issue     = {6},
  numpages  = {6},
  publisher = {American Physical Society},
  url       = {https://link.aps.org/doi/10.1103/PhysRevA.109.L061502},
}

@Article{Longhi2025,
  author    = {Longhi, Stefano},
  journal   = {Phys. Rev. Lett.},
  title     = {Erratic Non-Hermitian Skin Localization},
  year      = {2025},
  month     = {May},
  pages     = {196302},
  volume    = {134},
  doi       = {10.1103/PhysRevLett.134.196302},
  issue     = {19},
  numpages  = {7},
  publisher = {American Physical Society},
  url       = {https://link.aps.org/doi/10.1103/PhysRevLett.134.196302},
}

@Article{Burger1999,
  author    = {Burger, S. and Bongs, K. and Dettmer, S. and Ertmer, W. and Sengstock, K. and Sanpera, A. and Shlyapnikov, G. V. and Lewenstein, M.},
  journal   = {Phys. Rev. Lett.},
  title     = {Dark Solitons in Bose-Einstein Condensates},
  year      = {1999},
  month     = {Dec},
  pages     = {5198--5201},
  volume    = {83},
  doi       = {10.1103/PhysRevLett.83.5198},
  issue     = {25},
  numpages  = {0},
  publisher = {American Physical Society},
  url       = {https://link.aps.org/doi/10.1103/PhysRevLett.83.5198},
}

@Article{Dobrek1999,
  author    = {Dobrek, \L{}. and Gajda, M. and Lewenstein, M. and Sengstock, K. and Birkl, G. and Ertmer, W.},
  journal   = {Phys. Rev. A},
  title     = {Optical generation of vortices in trapped Bose-Einstein condensates},
  year      = {1999},
  month     = {Nov},
  pages     = {R3381--R3384},
  volume    = {60},
  doi       = {10.1103/PhysRevA.60.R3381},
  issue     = {5},
  numpages  = {0},
  publisher = {American Physical Society},
  url       = {https://link.aps.org/doi/10.1103/PhysRevA.60.R3381},
}

@Article{Denschlag2000,
  author    = {Denschlag, J. and Simsarian, J. E. and Feder, D. L. and Clark, Charles W. and Collins, L. A. and Cubizolles, J. and Deng, L. and Hagley, E. W. and Helmerson, K. and Reinhardt, W. P. and Rolston, S. L. and Schneider, B. I. and Phillips, W. D.},
  journal   = {Science},
  title     = {Generating Solitons by Phase Engineering of a Bose-Einstein Condensate},
  year      = {2000},
  issn      = {1095-9203},
  month     = jan,
  number    = {5450},
  pages     = {97--101},
  volume    = {287},
  doi       = {10.1126/science.287.5450.97},
  publisher = {American Association for the Advancement of Science (AAAS)},
}

@Article{Borgnia2020,
  author    = {Borgnia, Dan S. and Kruchkov, Alex Jura and Slager, Robert-Jan},
  journal   = {Phys. Rev. Lett.},
  title     = {Non-Hermitian Boundary Modes and Topology},
  year      = {2020},
  month     = {Feb},
  pages     = {056802},
  volume    = {124},
  doi       = {10.1103/PhysRevLett.124.056802},
  issue     = {5},
  numpages  = {6},
  publisher = {American Physical Society},
  url       = {https://link.aps.org/doi/10.1103/PhysRevLett.124.056802},
}

@Article{Okuma2020,
  author    = {Okuma, Nobuyuki and Kawabata, Kohei and Shiozaki, Ken and Sato, Masatoshi},
  journal   = {Phys. Rev. Lett.},
  title     = {Topological Origin of Non-Hermitian Skin Effects},
  year      = {2020},
  month     = {Feb},
  pages     = {086801},
  volume    = {124},
  doi       = {10.1103/PhysRevLett.124.086801},
  issue     = {8},
  numpages  = {7},
  publisher = {American Physical Society},
  url       = {https://link.aps.org/doi/10.1103/PhysRevLett.124.086801},
}

@Article{KZhang2020,
  author    = {Zhang, Kai and Yang, Zhesen and Fang, Chen},
  journal   = {Phys. Rev. Lett.},
  title     = {Correspondence between Winding Numbers and Skin Modes in Non-Hermitian Systems},
  year      = {2020},
  month     = {Sep},
  pages     = {126402},
  volume    = {125},
  doi       = {10.1103/PhysRevLett.125.126402},
  issue     = {12},
  numpages  = {6},
  publisher = {American Physical Society},
  url       = {https://link.aps.org/doi/10.1103/PhysRevLett.125.126402},
}

@Article{Evers2008RMP,
  author    = {Evers, Ferdinand and Mirlin, Alexander D.},
  journal   = {Rev. Mod. Phys.},
  title     = {Anderson transitions},
  year      = {2008},
  month     = {Oct},
  pages     = {1355--1417},
  volume    = {80},
  doi       = {10.1103/RevModPhys.80.1355},
  issue     = {4},
  numpages  = {0},
  publisher = {American Physical Society},
  url       = {https://link.aps.org/doi/10.1103/RevModPhys.80.1355},
}

@Article{Schehr2012,
  author    = {Schehr, Gr\'egory and Majumdar, Satya N.},
  journal   = {Phys. Rev. Lett.},
  title     = {Universal Order Statistics of Random Walks},
  year      = {2012},
  month     = {Jan},
  pages     = {040601},
  volume    = {108},
  doi       = {10.1103/PhysRevLett.108.040601},
  issue     = {4},
  numpages  = {5},
  publisher = {American Physical Society},
  url       = {https://link.aps.org/doi/10.1103/PhysRevLett.108.040601},
}

@Article{Majumdar2020,
  author    = {Majumdar, Satya N. and Pal, Arnab and Schehr, Grégory},
  journal   = {Phys. Rep.},
  title     = {Extreme value statistics of correlated random variables: A pedagogical review},
  year      = {2020},
  issn      = {0370-1573},
  month     = jan,
  pages     = {1--32},
  volume    = {840},
  doi       = {10.1016/j.physrep.2019.10.005},
  publisher = {Elsevier BV},
}

@Article{YWang2020,
  author    = {Wang, Yucheng and Zhang, Long and Niu, Sen and Yu, Dapeng and Liu, Xiong-Jun},
  journal   = {Phys. Rev. Lett.},
  title     = {Realization and Detection of Nonergodic Critical Phases in an Optical Raman Lattice},
  year      = {2020},
  month     = {Aug},
  pages     = {073204},
  volume    = {125},
  doi       = {10.1103/PhysRevLett.125.073204},
  issue     = {7},
  numpages  = {7},
  publisher = {American Physical Society},
  url       = {https://link.aps.org/doi/10.1103/PhysRevLett.125.073204},
}

@Article{Goncalves2024,
  author    = {Gonçalves, Miguel and Amorim, Bruno and Riche, Flavio and Castro, Eduardo V. and Ribeiro, Pedro},
  journal   = {Nat. Phys.},
  title     = {Incommensurability enabled quasi-fractal order in 1D narrow-band moiré systems},
  year      = {2024},
  issn      = {1745-2481},
  month     = oct,
  number    = {12},
  pages     = {1933--1940},
  volume    = {20},
  doi       = {10.1038/s41567-024-02662-2},
  publisher = {Springer Science and Business Media LLC},
}

@Article{Shimomura2024,
  author    = {Shimomura, Kenji and Sato, Masatoshi},
  journal   = {Phys. Rev. Lett.},
  title     = {General Criterion for Non-Hermitian Skin Effects and Application: Fock Space Skin Effects in Many-Body Systems},
  year      = {2024},
  month     = {Sep},
  pages     = {136502},
  volume    = {133},
  doi       = {10.1103/PhysRevLett.133.136502},
  issue     = {13},
  numpages  = {7},
  publisher = {American Physical Society},
  url       = {https://link.aps.org/doi/10.1103/PhysRevLett.133.136502},
}

@Article{Gliozzi2024,
  author    = {Gliozzi, Jacopo and De Tomasi, Giuseppe and Hughes, Taylor L.},
  journal   = {Phys. Rev. Lett.},
  title     = {Many-Body Non-Hermitian Skin Effect for Multipoles},
  year      = {2024},
  month     = {Sep},
  pages     = {136503},
  volume    = {133},
  doi       = {10.1103/PhysRevLett.133.136503},
  issue     = {13},
  numpages  = {7},
  publisher = {American Physical Society},
  url       = {https://link.aps.org/doi/10.1103/PhysRevLett.133.136503},
}

@Article{ZPang2024,
  author    = {Pang, Zehai and Wong, Bengy Tsz Tsun and Hu, Jinbing and Yang, Yi},
  journal   = {Phys. Rev. Lett.},
  title     = {Synthetic Non-Abelian Gauge Fields for Non-Hermitian Systems},
  year      = {2024},
  month     = {Jan},
  pages     = {043804},
  volume    = {132},
  doi       = {10.1103/PhysRevLett.132.043804},
  issue     = {4},
  numpages  = {7},
  publisher = {American Physical Society},
  url       = {https://link.aps.org/doi/10.1103/PhysRevLett.132.043804},
}

@Article{RJChen2025,
  author    = {Chen, Rui-Jie and Zhang, Guo-Qing and Li, Zhi and Zhang, Dan-Wei},
  journal   = {Phys. Rev. A},
  title     = {Mobility rings in a non-Hermitian non-Abelian quasiperiodic lattice},
  year      = {2025},
  month     = {Jul},
  pages     = {013320},
  volume    = {112},
  doi       = {10.1103/zfrn-53lz},
  issue     = {1},
  numpages  = {8},
  publisher = {American Physical Society},
  url       = {https://link.aps.org/doi/10.1103/zfrn-53lz},
}

@Article{Sanahal2025,
  author    = {Sanahal, Moirangthem and Panda, Subhasis and Nandy, Snehasish},
  journal   = {Phys. Rev. B},
  title     = {Gauge field induced skin effect in spinful non-Hermitian systems},
  year      = {2025},
  month     = {Sep},
  pages     = {125149},
  volume    = {112},
  doi       = {10.1103/vhz9-xwf4},
  issue     = {12},
  numpages  = {7},
  publisher = {American Physical Society},
  url       = {https://link.aps.org/doi/10.1103/vhz9-xwf4},
}

@Article{WZhang2025,
  author    = {Zhang, Wenna and Hu, Yutao and Zhang, Hongyi and Liu, Xiang and Shen, Yuecheng and Veronis, Georgios and Al\`u, Andrea and Huang, Yin and Luo, Wenchen},
  journal   = {Phys. Rev. B},
  title     = {Skin effect in non-Hermitian systems with SU(2) gauge fields},
  year      = {2025},
  month     = {Sep},
  pages     = {125164},
  volume    = {112},
  doi       = {10.1103/cqmg-qzt4},
  issue     = {12},
  numpages  = {16},
  publisher = {American Physical Society},
  url       = {https://link.aps.org/doi/10.1103/cqmg-qzt4},
}

@Article{Yoshida2024,
  author    = {Yoshida, Tsuneya and Zhang, Song-Bo and Neupert, Titus and Kawakami, Norio},
  journal   = {Phys. Rev. Lett.},
  title     = {Non-Hermitian Mott Skin Effect},
  year      = {2024},
  month     = {Aug},
  pages     = {076502},
  volume    = {133},
  doi       = {10.1103/PhysRevLett.133.076502},
  issue     = {7},
  numpages  = {8},
  publisher = {American Physical Society},
  url       = {https://link.aps.org/doi/10.1103/PhysRevLett.133.076502},
}

@Article{YQin2025,
  author    = {Qin, Yi and Ang, Yee Sin and Lee, Ching Hua and Li, Linhu},
  journal   = {Commun. Phys.},
  title     = {Many-body critical non-Hermitian skin effect},
  year      = {2025},
  issn      = {2399-3650},
  month     = dec,
  number    = {1},
  volume    = {9},
  doi       = {10.1038/s42005-025-02448-9},
  publisher = {Springer Science and Business Media LLC},
}

@Article{XChen2026,
  author    = {Chen, Xiangru and Wu, Jien and Chen, Xingyu and Pu, Zhenhang and Hu, Yejian and Lu, Jiuyang and Ke, Manzhu and Deng, Weiyin and Liu, Zhengyou},
  journal   = {Phys. Rev. Lett.},
  title     = {Implementing Non-Abelian Hatano-Nelson Model in Electric Circuits},
  year      = {2026},
  month     = {Mar},
  pages     = {106604},
  volume    = {136},
  doi       = {10.1103/48wx-5gmj},
  issue     = {10},
  numpages  = {6},
  publisher = {American Physical Society},
  url       = {https://link.aps.org/doi/10.1103/48wx-5gmj},
}

@Article{YMiao2025,
  author    = {Miao, Yazhuang and Zhao, Yiming and Wang, Yong and Qiao, Jie and Zhao, Xiaolong and Yi, Xuexi},
  journal   = {Phys. Rev. A},
  title     = {Non-Abelian gauge enhances self-healing for non-Hermitian Su-Schrieffer-Heeger chain},
  year      = {2025},
  month     = {Nov},
  pages     = {053302},
  volume    = {112},
  doi       = {10.1103/148x-k5xd},
  issue     = {5},
  numpages  = {9},
  publisher = {American Physical Society},
  url       = {https://link.aps.org/doi/10.1103/148x-k5xd},
}

@Article{JXZhong2026,
  author        = {Zhong, Jia-Xin and Kim, Jee Woo and Longhi, Stefano and Jing, Yun},
  journal       = {arXiv},
  title         = {Observation of Erratic Non-Hermitian Skin Localization and Transport},
  year          = {2026},
  month         = jan,
  archiveprefix = {arXiv},
  copyright     = {arXiv.org perpetual, non-exclusive license},
  eprint        = {2601.19749},
  primaryclass  = {cond-mat.dis-nn},
  publisher     = {arXiv},
}

@Article{Jagannathan2021,
  author    = {Jagannathan, Anuradha},
  journal   = {Rev. Mod. Phys.},
  title     = {The Fibonacci quasicrystal: Case study of hidden dimensions and multifractality},
  year      = {2021},
  month     = {Nov},
  pages     = {045001},
  volume    = {93},
  doi       = {10.1103/RevModPhys.93.045001},
  issue     = {4},
  numpages  = {37},
  publisher = {American Physical Society},
  url       = {https://link.aps.org/doi/10.1103/RevModPhys.93.045001},
}

@Article{Gefen1980,
  author    = {Gefen, Yuval and Mandelbrot, Benoit B. and Aharony, Amnon},
  journal   = {Phys. Rev. Lett.},
  title     = {Critical Phenomena on Fractal Lattices},
  year      = {1980},
  month     = {Sep},
  pages     = {855--858},
  volume    = {45},
  doi       = {10.1103/PhysRevLett.45.855},
  issue     = {11},
  numpages  = {0},
  publisher = {American Physical Society},
  url       = {https://link.aps.org/doi/10.1103/PhysRevLett.45.855},
}

@Article{Nuckolls2025,
  author    = {Nuckolls, Kevin P. and Scheer, Michael G. and Wong, Dillon and Oh, Myungchul and Lee, Ryan L. and Herzog-Arbeitman, Jonah and Watanabe, Kenji and Taniguchi, Takashi and Lian, Biao and Yazdani, Ali},
  journal   = {Nature},
  title     = {Spectroscopy of the fractal Hofstadter energy spectrum},
  year      = {2025},
  issn      = {1476-4687},
  month     = feb,
  number    = {8053},
  pages     = {60--66},
  volume    = {639},
  doi       = {10.1038/s41586-024-08550-2},
  publisher = {Springer Science and Business Media LLC},
}

@Article{Hofstadter1976,
  author    = {Hofstadter, Douglas R.},
  journal   = {Phys. Rev. B},
  title     = {Energy levels and wave functions of Bloch electrons in rational and irrational magnetic fields},
  year      = {1976},
  month     = {Sep},
  pages     = {2239--2249},
  volume    = {14},
  doi       = {10.1103/PhysRevB.14.2239},
  issue     = {6},
  numpages  = {0},
  publisher = {American Physical Society},
  url       = {https://link.aps.org/doi/10.1103/PhysRevB.14.2239},
}

@Article{XTWang2026,
  author    = {Wan, Xu-Tao and Gao, Chao and Shi, Zhe-Yu},
  journal   = {Phys. Rev. Lett.},
  title     = {Fractal Spectrum in Twisted Bilayer Optical Lattice},
  year      = {2026},
  month     = {Jan},
  pages     = {033401},
  volume    = {136},
  doi       = {10.1103/nz52-ypsw},
  issue     = {3},
  numpages  = {8},
  publisher = {American Physical Society},
  url       = {https://link.aps.org/doi/10.1103/nz52-ypsw},
}

@Article{PWang2019,
  author    = {Wang, Peng and Zheng, Yuanlin and Chen, Xianfeng and Huang, Changming and Kartashov, Yaroslav V. and Torner, Lluis and Konotop, Vladimir V. and Ye, Fangwei},
  journal   = {Nature},
  title     = {Localization and delocalization of light in photonic moiré lattices},
  year      = {2019},
  issn      = {1476-4687},
  month     = dec,
  number    = {7788},
  pages     = {42--46},
  volume    = {577},
  doi       = {10.1038/s41586-019-1851-6},
  publisher = {Springer Science and Business Media LLC},
}

@Article{ZMeng2023,
  author    = {Meng, Zengming and Wang, Liangwei and Han, Wei and Liu, Fangde and Wen, Kai and Gao, Chao and Wang, Pengjun and Chin, Cheng and Zhang, Jing},
  journal   = {Nature},
  title     = {Atomic Bose–Einstein condensate in twisted-bilayer optical lattices},
  year      = {2023},
  issn      = {1476-4687},
  month     = feb,
  number    = {7951},
  pages     = {231--236},
  volume    = {615},
  doi       = {10.1038/s41586-023-05695-4},
  publisher = {Springer Science and Business Media LLC},
}

@Article{JHan1994,
  author    = {Han, J. H. and Thouless, D. J. and Hiramoto, H. and Kohmoto, M.},
  journal   = {Phys. Rev. B},
  title     = {Critical and bicritical properties of Harper's equation with next-nearest-neighbor coupling},
  year      = {1994},
  month     = {Oct},
  pages     = {11365--11380},
  volume    = {50},
  doi       = {10.1103/PhysRevB.50.11365},
  issue     = {16},
  numpages  = {0},
  publisher = {American Physical Society},
  url       = {https://link.aps.org/doi/10.1103/PhysRevB.50.11365},
}

@Article{Chang1997,
  author    = {Chang, I. and Ikezawa, K. and Kohmoto, M.},
  journal   = {Phys. Rev. B},
  title     = {Multifractal properties of the wave functions of the square-lattice tight-binding model with next-nearest-neighbor hopping in a magnetic field},
  year      = {1997},
  month     = {May},
  pages     = {12971--12975},
  volume    = {55},
  doi       = {10.1103/PhysRevB.55.12971},
  issue     = {19},
  numpages  = {0},
  publisher = {American Physical Society},
  url       = {https://link.aps.org/doi/10.1103/PhysRevB.55.12971},
}

@Article{FLiu2015,
  author    = {Liu, Fangli and Ghosh, Somnath and Chong, Y. D.},
  journal   = {Phys. Rev. B},
  title     = {Localization and adiabatic pumping in a generalized Aubry-Andr\'e-Harper model},
  year      = {2015},
  month     = {Jan},
  pages     = {014108},
  volume    = {91},
  doi       = {10.1103/PhysRevB.91.014108},
  issue     = {1},
  numpages  = {9},
  publisher = {American Physical Society},
  url       = {https://link.aps.org/doi/10.1103/PhysRevB.91.014108},
}

@Article{JWang2016,
  author    = {Wang, Jun and Liu, Xia-Ji and Xianlong, Gao and Hu, Hui},
  journal   = {Phys. Rev. B},
  title     = {Phase diagram of a non-Abelian Aubry-Andr\'e-Harper model with $p$-wave superfluidity},
  year      = {2016},
  month     = {Mar},
  pages     = {104504},
  volume    = {93},
  doi       = {10.1103/PhysRevB.93.104504},
  issue     = {10},
  numpages  = {7},
  publisher = {American Physical Society},
  url       = {https://link.aps.org/doi/10.1103/PhysRevB.93.104504},
}

@Article{XJLiu2024,
  author    = {Liu, Xiong-Jun},
  journal   = {Nat. Phys.},
  title     = {Quantum matter in multifractal patterns},
  year      = {2024},
  issn      = {1745-2481},
  month     = oct,
  number    = {12},
  pages     = {1851--1852},
  volume    = {20},
  doi       = {10.1038/s41567-024-02663-1},
  publisher = {Springer Science and Business Media LLC},
}

@Article{YWang2022,
  author    = {Wang, Yucheng and Zhang, Long and Sun, Wei and Poon, Ting-Fung Jeffrey and Liu, Xiong-Jun},
  journal   = {Phys. Rev. B},
  title     = {Quantum phase with coexisting localized, extended, and critical zones},
  year      = {2022},
  month     = {Oct},
  pages     = {L140203},
  volume    = {106},
  doi       = {10.1103/PhysRevB.106.L140203},
  issue     = {14},
  numpages  = {7},
  publisher = {American Physical Society},
  url       = {https://link.aps.org/doi/10.1103/PhysRevB.106.L140203},
}

@Article{HYao2019,
  author    = {Yao, Hepeng and Khoudli, Alice and Bresque, L\'ea and Sanchez-Palencia, Laurent},
  journal   = {Phys. Rev. Lett.},
  title     = {Critical Behavior and Fractality in Shallow One-Dimensional Quasiperiodic Potentials},
  year      = {2019},
  month     = {Aug},
  pages     = {070405},
  volume    = {123},
  doi       = {10.1103/PhysRevLett.123.070405},
  issue     = {7},
  numpages  = {6},
  publisher = {American Physical Society},
  url       = {https://link.aps.org/doi/10.1103/PhysRevLett.123.070405},
}

@Article{Migue2023b,
  author    = {Gon\ifmmode \mbox{\c{c}}\else \c{c}\fi{}alves, Miguel and Amorim, B. and Castro, Eduardo V. and Ribeiro, Pedro},
  journal   = {Phys. Rev. B},
  title     = {Renormalization group theory of one-dimensional quasiperiodic lattice models with commensurate approximants},
  year      = {2023},
  month     = {Sep},
  pages     = {L100201},
  volume    = {108},
  doi       = {10.1103/PhysRevB.108.L100201},
  issue     = {10},
  numpages  = {6},
  publisher = {American Physical Society},
  url       = {https://link.aps.org/doi/10.1103/PhysRevB.108.L100201},
}

@Article{Banerjee2025,
  author    = {Banerjee, Sanchayan and Padhi, Soumya Ranjan and Mishra, Tapan},
  journal   = {Phys. Rev. B},
  title     = {Emergence of distinct exact mobility edges in a quasiperiodic chain},
  year      = {2025},
  month     = {Jun},
  pages     = {L220201},
  volume    = {111},
  doi       = {10.1103/PhysRevB.111.L220201},
  issue     = {22},
  numpages  = {7},
  publisher = {American Physical Society},
  url       = {https://link.aps.org/doi/10.1103/PhysRevB.111.L220201},
}

@Article{Rispoli2019,
  author    = {Rispoli, Matthew and Lukin, Alexander and Schittko, Robert and Kim, Sooshin and Tai, M. Eric and Léonard, Julian and Greiner, Markus},
  journal   = {Nature},
  title     = {Quantum critical behaviour at the many-body localization transition},
  year      = {2019},
  issn      = {1476-4687},
  month     = sep,
  number    = {7774},
  pages     = {385--389},
  volume    = {573},
  doi       = {10.1038/s41586-019-1527-2},
  publisher = {Springer Science and Business Media LLC},
}

@Article{TXiao2021,
  author    = {Xiao, Teng and Xie, Dizhou and Dong, Zhaoli and Chen, Tao and Yi, Wei and Yan, Bo},
  journal   = {Sci. Bull.},
  title     = {Observation of topological phase with critical localization in a quasi-periodic lattice},
  year      = {2021},
  issn      = {2095-9273},
  month     = nov,
  number    = {21},
  pages     = {2175--2180},
  volume    = {66},
  doi       = {10.1016/j.scib.2021.07.025},
  publisher = {Elsevier BV},
}

@Article{HLi2023,
  author    = {Li, Hao and Wang, Yong-Yi and Shi, Yun-Hao and Huang, Kaixuan and Song, Xiaohui and Liang, Gui-Han and Mei, Zheng-Yang and Zhou, Bozhen and Zhang, He and Zhang, Jia-Chi and Chen, Shu and Zhao, S. P. and Tian, Ye and Yang, Zhan-Ying and Xiang, Zhongcheng and Xu, Kai and Zheng, Dongning and Fan, Heng},
  journal   = {npj Quantum Inf.},
  title     = {Observation of critical phase transition in a generalized Aubry-André-Harper model with superconducting circuits},
  year      = {2023},
  issn      = {2056-6387},
  month     = apr,
  number    = {1},
  volume    = {9},
  doi       = {10.1038/s41534-023-00712-w},
  publisher = {Springer Science and Business Media LLC},
}

@Article{Shimasaki2024,
  author    = {Shimasaki, Toshihiko and Prichard, Max and Kondakci, H. Esat and Pagett, Jared E. and Bai, Yifei and Dotti, Peter and Cao, Alec and Dardia, Anna R. and Lu, Tsung-Cheng and Grover, Tarun and Weld, David M.},
  journal   = {Nat. Phys.},
  title     = {Anomalous localization in a kicked quasicrystal},
  year      = {2024},
  issn      = {1745-2481},
  month     = jan,
  number    = {3},
  pages     = {409--414},
  volume    = {20},
  doi       = {10.1038/s41567-023-02329-4},
  publisher = {Springer Science and Business Media LLC},
}

@Article{WHuang2025,
  author        = {Huang, Wenhui and Zhou, Xin-Chi and Zhang, Libo and Zhang, Jiawei and Zhou, Yuxuan and Yao, Bing-Chen and Guo, Zechen and Huang, Peisheng and Li, Qixian and Liang, Yongqi and Liu, Yiting and Qiu, Jiawei and Sun, Daxiong and Sun, Xuandong and Wang, Zilin and Xie, Changrong and Xiong, Yuzhe and Yang, Xiaohan and Zhang, Jiajian and Zhang, Zihao and Chu, Ji and Guo, Weijie and Jiang, Ji and Linpeng, Xiayu and Ren, Wenhui and Yuan, Yuefeng and Niu, Jingjing and Tao, Ziyu and Liu, Song and Zhong, Youpeng and Liu, Xiong-Jun and Yu, Dapeng},
  journal       = {arXiv},
  title         = {Experimental observation of exact quantum critical states},
  year          = {2025},
  month         = feb,
  archiveprefix = {arXiv},
  copyright     = {arXiv.org perpetual, non-exclusive license},
  eprint        = {2502.19185},
  publisher     = {arXiv},
}

@Article{Lee2016,
  author    = {Lee, Tony E.},
  journal   = {Phys. Rev. Lett.},
  title     = {Anomalous Edge State in a Non-Hermitian Lattice},
  year      = {2016},
  month     = {Apr},
  pages     = {133903},
  volume    = {116},
  doi       = {10.1103/PhysRevLett.116.133903},
  issue     = {13},
  numpages  = {5},
  publisher = {American Physical Society},
  url       = {https://link.aps.org/doi/10.1103/PhysRevLett.116.133903},
}

@Article{QYao2024,
  author    = {Yao, Qi and Yang, Xiaotian and Iliasov, Askar A. and Katsnelson, Mikhail I. and Yuan, Shengjun},
  journal   = {Phys. Rev. B},
  title     = {Wave functions in the critical phase: A planar Sierpi\ifmmode \acute{n}\else \'{n}\fi{}ski fractal lattice},
  year      = {2024},
  month     = {Jul},
  pages     = {035403},
  volume    = {110},
  doi       = {10.1103/PhysRevB.110.035403},
  issue     = {3},
  numpages  = {9},
  publisher = {American Physical Society},
  url       = {https://link.aps.org/doi/10.1103/PhysRevB.110.035403},
}

@Article{JSun2024,
  author    = {Sun, Junsong and Li, Chang-An and Guo, Qingyang and Zhang, Weixuan and Feng, Shiping and Zhang, Xiangdong and Guo, Huaiming and Trauzettel, Bj\"orn},
  journal   = {Phys. Rev. B},
  title     = {Non-Hermitian quantum fractals},
  year      = {2024},
  month     = {Nov},
  pages     = {L201103},
  volume    = {110},
  doi       = {10.1103/PhysRevB.110.L201103},
  issue     = {20},
  numpages  = {6},
  publisher = {American Physical Society},
  url       = {https://link.aps.org/doi/10.1103/PhysRevB.110.L201103},
}

@Article{XCai2022,
  author    = {Cai, Xiaoming},
  journal   = {Phys. Rev. B},
  title     = {Localization transitions and winding numbers for non-Hermitian Aubry-Andr\'e-Harper models with off-diagonal modulations},
  year      = {2022},
  month     = {Dec},
  pages     = {214207},
  volume    = {106},
  doi       = {10.1103/PhysRevB.106.214207},
  issue     = {21},
  numpages  = {9},
  publisher = {American Physical Society},
  url       = {https://link.aps.org/doi/10.1103/PhysRevB.106.214207},
}

@Article{HLiang2025,
  author        = {Liang, Hui-Qiang and Li, Linhu and Xu, Guo-Fu},
  journal       = {arXiv},
  title         = {Size-dependent critical localization},
  year          = {2025},
  month         = sep,
  archiveprefix = {arXiv},
  copyright     = {Creative Commons Attribution 4.0 International},
  eprint        = {2509.18943},
  primaryclass  = {cond-mat.dis-nn},
  publisher     = {arXiv},
}

@Article{Claes2021,
  author    = {Claes, Jahan and Hughes, Taylor L.},
  journal   = {Phys. Rev. B},
  title     = {Skin effect and winding number in disordered non-Hermitian systems},
  year      = {2021},
  month     = {Apr},
  pages     = {L140201},
  volume    = {103},
  doi       = {10.1103/PhysRevB.103.L140201},
  issue     = {14},
  numpages  = {7},
  publisher = {American Physical Society},
  url       = {https://link.aps.org/doi/10.1103/PhysRevB.103.L140201},
}

@Article{ZQZhang2023,
  author    = {Zhang, Zhi-Qiang and Liu, Hongfang and Liu, Haiwen and Jiang, Hua and Xie, X.C.},
  journal   = {Sci. Bull.},
  title     = {Bulk-boundary correspondence in disordered non-Hermitian systems},
  year      = {2023},
  issn      = {2095-9273},
  month     = jan,
  number    = {2},
  pages     = {157--164},
  volume    = {68},
  doi       = {10.1016/j.scib.2023.01.002},
  publisher = {Elsevier BV},
}

@Article{HNie2026,
  author        = {Nie, Haoran and Jiang, Chaoran and Shen, Xiangying and Xu, Lei},
  journal       = {arXiv},
  title         = {Imaginary Gauge Field and Non-Hermitian Topological Transition Emerging Through Attenuation-Gauge Duality in Conservative Systems},
  year          = {2026},
  month         = mar,
  archiveprefix = {arXiv},
  copyright     = {arXiv.org perpetual, non-exclusive license},
  eprint        = {2603.17557},
  primaryclass  = {cond-mat.other},
  publisher     = {arXiv},
}

@Article{JTao2026,
  author    = {Tao, J. and Mercado-Gutierrez, E. D. and Zhao, M. and Spielman, I. B.},
  journal   = {Phys. Rev. Lett.},
  title     = {Imaginary Gauge Potentials in a Non-Hermitian Spin-Orbit Coupled Quantum Gas},
  year      = {2026},
  month     = {Mar},
  pages     = {113401},
  volume    = {136},
  doi       = {10.1103/tpfc-n3bq},
  issue     = {11},
  numpages  = {7},
  publisher = {American Physical Society},
  url       = {https://link.aps.org/doi/10.1103/tpfc-n3bq},
}

@Article{Helbig2020,
  author    = {Helbig, T. and Hofmann, T. and Imhof, S. and Abdelghany, M. and Kiessling, T. and Molenkamp, L. W. and Lee, C. H. and Szameit, A. and Greiter, M. and Thomale, R.},
  journal   = {Nat. Phys.},
  title     = {Generalized bulk–boundary correspondence in non-Hermitian topolectrical circuits},
  year      = {2020},
  issn      = {1745-2481},
  month     = jun,
  number    = {7},
  pages     = {747--750},
  volume    = {16},
  doi       = {10.1038/s41567-020-0922-9},
  publisher = {Springer Science and Business Media LLC},
}

@Article{Weidemann2020,
  author    = {Weidemann, Sebastian and Kremer, Mark and Helbig, Tobias and Hofmann, Tobias and Stegmaier, Alexander and Greiter, Martin and Thomale, Ronny and Szameit, Alexander},
  journal   = {Science},
  title     = {Topological funneling of light},
  year      = {2020},
  issn      = {1095-9203},
  month     = apr,
  number    = {6488},
  pages     = {311--314},
  volume    = {368},
  doi       = {10.1126/science.aaz8727},
  publisher = {American Association for the Advancement of Science (AAAS)},
}

@Article{SLi2026,
  author    = {Li, Shan-Zhong and Zhang, Yi-Cai and Wang, Yucheng and Zhang, Shanchao and Zhu, Shi-Liang and Li, Zhi},
  journal   = {Sci. China-Phys. Mech. Astron.},
  title     = {Multifractal-enriched mobility edges and emergent quantum phases in Rydberg atomic arrays},
  year      = {2026},
  issn      = {1869-1927},
  month     = sep,
  number    = {1},
  volume    = {69},
  doi       = {10.1007/s11433-025-2774-2},
  publisher = {Springer Science and Business Media LLC},
}

@Article{Han2026,
  author    = {Han, Yu-Hong and Li, Yi and Zhang, Jia-Hui and Kou, Yang and Xiao, Liantuan and Jia, Suotang and Li, Linhu and Mei, Feng},
  journal   = {Chin. Phys. Lett.},
  title     = {Observation of gauge field induced non-Hermitian helical skin effects},
  year      = {2026},
  issn      = {1741-3540},
  month     = feb,
  doi       = {10.1088/0256-307x/43/4/040704},
  publisher = {IOP Publishing},
}

@Article{ZChen2026,
  author        = {Chen, Zhangyuan and Idrees, Muhammad and Yang, Ying and Tong, Xianqi and Yang, Xiaosen},
  journal       = {arXiv},
  title         = {Quasiperiodic Skin Criticality in an Exactly Solvable Non-Hermitian Quasicrystal},
  year          = {2026},
  month         = jan,
  archiveprefix = {arXiv},
  copyright     = {arXiv.org perpetual, non-exclusive license},
  eprint        = {2601.23015},
  primaryclass  = {cond-mat.mes-hall},
  publisher     = {arXiv},
}

@Article{XZhou2026,
  author    = {Zhou, Xin-Chi and Yao, Bing-Chen and Wang, Yongjian and Wang, Yucheng and Wei, Yudong and Zhou, Qi and Liu, Xiong-Jun},
  journal   = {Sci. Bull.},
  title     = {The fundamental localization phases in quasiperiodic systems: a unified framework and exact results},
  year      = {2026},
  issn      = {2095-9273},
  month     = mar,
  doi       = {10.1016/j.scib.2026.03.002},
  publisher = {Elsevier BV},
}

@Misc{Nan2026,
  author    = {Nan, Guolin and Li, Zhijian and Mei, Feng and Xu, Zhihao},
  title     = {Anomalous Localization and Mobility Edges in Non-Hermitian Quasicrystals with Disordered Imaginary Gauge Fields},
  year      = {2026},
  copyright = {Creative Commons Attribution 4.0 International},
  doi       = {10.48550/ARXIV.2601.14754},
  publisher = {arXiv},
}

@Article{Chakrabarty2025,
  author    = {Chakrabarty, Aditi and Datta, Sanjoy},
  journal   = {Phys. Rev. B},
  title     = {Fate of Wannier-Stark localization and skin effect in periodically driven non-Hermitian quasiperiodic lattices},
  year      = {2025},
  month     = {May},
  pages     = {174202},
  volume    = {111},
  doi       = {10.1103/PhysRevB.111.174202},
  issue     = {17},
  numpages  = {15},
  publisher = {American Physical Society},
  url       = {https://link.aps.org/doi/10.1103/PhysRevB.111.174202},
}

@Article{YPWang2025,
  author    = {Wang, Yu-Peng and Chang, Chuo-Kai and Okugawa, Ryo and Hsu, Chen-Hsuan},
  journal   = {Phys. Rev. Res.},
  title     = {Quasiperiodicity-induced bulk localization with self-similarity in non-Hermitian systems},
  year      = {2025},
  month     = {Dec},
  pages     = {043353},
  volume    = {7},
  doi       = {10.1103/8txw-m2cp},
  issue     = {4},
  numpages  = {14},
  publisher = {American Physical Society},
  url       = {https://link.aps.org/doi/10.1103/8txw-m2cp},
}

@Article{Chakrabarty2025b,
  author        = {Chakrabarty, Aditi and Banerjee, Sanchayan and Mishra, Tapan and Datta, Sanjoy},
  journal       = {arXiv},
  title         = {Emergence of non-trivial phases in interacting non-Hermitian quasiperiodic chains with power-law hopping},
  year          = {2025},
  month         = aug,
  archiveprefix = {arXiv},
  copyright     = {Creative Commons Attribution 4.0 International},
  eprint        = {2508.14724},
  primaryclass  = {cond-mat.dis-nn},
  publisher     = {arXiv},
}

@Article{Richardella2010,
  author    = {Richardella, Anthony and Roushan, Pedram and Mack, Shawn and Zhou, Brian and Huse, David A. and Awschalom, David D. and Yazdani, Ali},
  journal   = {Science},
  title     = {Visualizing Critical Correlations Near the Metal-Insulator Transition in Ga 1- x Mn x As},
  year      = {2010},
  issn      = {1095-9203},
  month     = feb,
  number    = {5966},
  pages     = {665--669},
  volume    = {327},
  doi       = {10.1126/science.1183640},
  publisher = {American Association for the Advancement of Science (AAAS)},
}

@Article{Biesenthal2022,
  author    = {Biesenthal, Tobias and Maczewsky, Lukas J. and Yang, Zhaoju and Kremer, Mark and Segev, Mordechai and Szameit, Alexander and Heinrich, Matthias},
  journal   = {Science},
  title     = {Fractal photonic topological insulators},
  year      = {2022},
  issn      = {1095-9203},
  month     = june,
  number    = {6597},
  pages     = {1114--1119},
  volume    = {376},
  doi       = {10.1126/science.abm2842},
  publisher = {American Association for the Advancement of Science (AAAS)},
}

@Article{XYXu2021,
  author    = {Xu, Xiao-Yun and Wang, Xiao-Wei and Chen, Dan-Yang and Smith, C. Morais and Jin, Xian-Min},
  journal   = {Nat. Photonics},
  title     = {Quantum transport in fractal networks},
  year      = {2021},
  issn      = {1749-4893},
  month     = july,
  number    = {9},
  pages     = {703--710},
  volume    = {15},
  doi       = {10.1038/s41566-021-00845-4},
  publisher = {Springer Science and Business Media LLC},
}

@Article{RojoFrancas2024,
  author    = {Rojo-Francàs, Abel and Pansari, Priyanshu and Bhattacharya, Utso and Juliá-Díaz, Bruno and Grass, Tobias},
  journal   = {Communications Physics},
  title     = {Anomalous quantum transport in fractal lattices},
  year      = {2024},
  issn      = {2399-3650},
  month     = aug,
  number    = {1},
  volume    = {7},
  doi       = {10.1038/s42005-024-01747-x},
  publisher = {Springer Science and Business Media LLC},
}

@Article{Ren2025,
  author    = {Ren, Wei and Zhu, Ziyan and Zhang, Xi and Luskin, Mitchell and Wang, Ke},
  journal   = {J. Phys. Condens. Matter},
  title     = {Review: moiré-of-moiré superlattice in twisted trilayer graphene},
  year      = {2025},
  issn      = {1361-648X},
  month     = Aug,
  number    = {35},
  pages     = {353001},
  volume    = {37},
  doi       = {10.1088/1361-648x/adf6f9},
  fjournal  = {Journal of Physics: Condensed Matter},
  publisher = {IOP Publishing},
}

@Article{Wang2019,
  author    = {Wang, Lujun and Zihlmann, Simon and Liu, Ming-Hao and Makk, Péter and Watanabe, Kenji and Taniguchi, Takashi and Baumgartner, Andreas and Schönenberger, Christian},
  journal   = {Nano Lett.},
  title     = {New Generation of Moiré Superlattices in Doubly Aligned hBN/Graphene/hBN Heterostructures},
  year      = {2019},
  issn      = {1530-6992},
  month     = Feb,
  number    = {4},
  pages     = {2371--2376},
  volume    = {19},
  doi       = {10.1021/acs.nanolett.8b05061},
  fjournal  = {Nano Letters},
  publisher = {American Chemical Society (ACS)},
}

@Article{Wang2024,
  author    = {Wang, Ce and Gao, Chao and Zhang, Jing and Zhai, Hui and Shi, Zhe-Yu},
  journal   = {Phys. Rev. Lett.},
  title     = {Three-Dimensional Moiré Crystal in Ultracold Atomic Gases},
  year      = {2024},
  issn      = {1079-7114},
  month     = Oct,
  number    = {16},
  pages     = {163401},
  volume    = {133},
  doi       = {10.1103/physrevlett.133.163401},
  publisher = {American Physical Society (APS)},
}

@Article{Wang2025,
  author        = {Wang, Ce and Gao, Chao and Shi, Zhe-Yu},
  journal       = {arXiv},
  title         = {Quasi-periodic moiré patterns and dimensional localization in three-dimensional quasi-moiré crystals},
  year          = {2025},
  month         = apr,
  archiveprefix = {arXiv},
  copyright     = {arXiv.org perpetual, non-exclusive license},
  eprint        = {2504.02574},
  primaryclass  = {cond-mat.quant-gas},
  publisher     = {arXiv},
}

@Misc{SM,
  author = {{Supplementary Materials}},
  note   = {See Supplementary Materials},
  year   = {2026},
}

@Article{WWang2022,
  author    = {Wang, Wei and Wang, Xulong and Ma, Guancong},
  journal   = {Nature},
  title     = {Non-Hermitian morphing of topological modes},
  year      = {2022},
  issn      = {1476-4687},
  month     = Aug,
  number    = {7921},
  pages     = {50--55},
  volume    = {608},
  doi       = {10.1038/s41586-022-04929-1},
  publisher = {Springer Science and Business Media LLC},
}

@Article{ZLin2026,
  author    = {Lin, Zhiyuan and Li, Jian and Song, Wange and Li, Xueyun and Xin, Haoran and Long, Xian and Chen, Chen and Zhu, Shining and Li, Tao and Zhang, Shuang},
  journal   = {Nat. Commun.},
  title     = {Artificial gauge fields for sculpting topological modes on photonic chips},
  year      = {2026},
  issn      = {2041-1723},
  month     = Apr,
  number    = {1},
  volume    = {17},
  doi       = {10.1038/s41467-026-71402-2},
  publisher = {Springer Science and Business Media LLC},
}

@Article{JWu2025,
  author    = {Wu, Jien and Hu, Yejian and He, Zhaojian and Deng, Ke and Huang, Xueqin and Ke, Manzhu and Deng, Weiyin and Lu, Jiuyang and Liu, Zhengyou},
  journal   = {Phys. Rev. Lett.},
  title     = {Hybrid-Order Skin Effect from Loss-Induced Nonreciprocity},
  year      = {2025},
  month     = {Apr},
  pages     = {176601},
  volume    = {134},
  doi       = {10.1103/PhysRevLett.134.176601},
  issue     = {17},
  numpages  = {7},
  publisher = {American Physical Society},
  url       = {https://link.aps.org/doi/10.1103/PhysRevLett.134.176601},
}


\vspace{3mm}
\begin{center}
  \textbf{\large End Matter}
\end{center}

\twocolumngrid

\begin{figure}[t]
    \centering
     \includegraphics[width=0.95\linewidth]{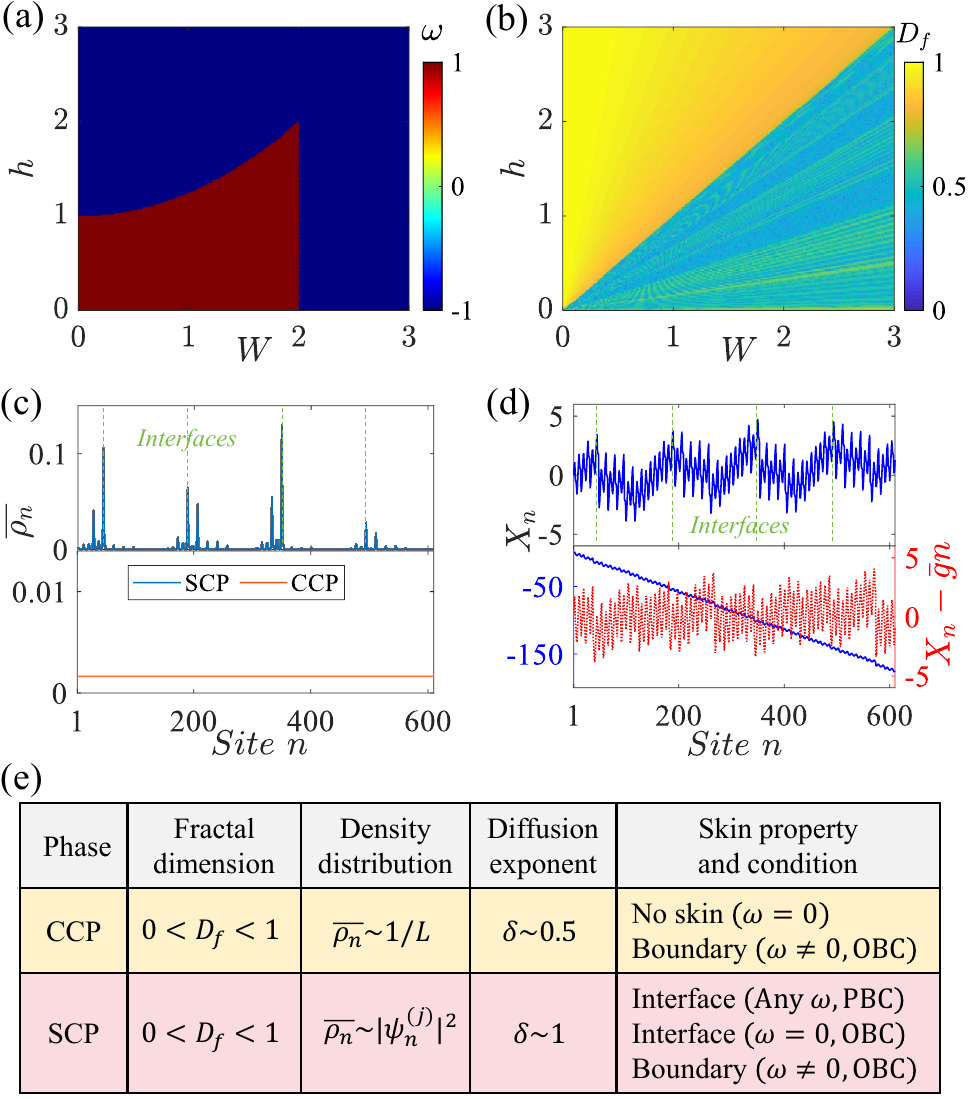}
     \caption{(a) Spectral winding number $\omega$ and (b) average fractal dimension $D_f$ as functions of $W$ and $h$. (c) Average density distribution $\overline{\rho_n}$ under the PBC for the SCP ($W=2$ and $h=1$) and the CCP (critical point of the AA model). (d) Corresponding cumulative imaginary phase $X_n$ (blue lines) and $X_n-\bar{g}n$ (red line) with quasiperiodic fluctuations and interfaces (green dash lines). (e) Summary of characteristic properties of the CCP and SCP.}
     \label{Efig1}
\end{figure}

{\it Properties of 1D SCP.—}In this part, we provide more results on topological, localization, and skin properties of the SCP in 1D. In this case, the exact wavefunction of the $j$-th eigenstate is given by  
\begin{equation}\label{1Dwavefunction}
    \psi_n^{(j)}=
    \begin{cases}
        \sin\left({\pi jn}/{(L+1)}\right)\exp\left(X_n\right), \quad \text{OBC}; 
        \\
         \exp\left(i{2\pi j n}/{L}-\bar{g}n+X_n\right),\quad \text{PBC},
    \end{cases}
\end{equation}
up to a normalization constant. The eigenenergies under the OBC and the PBC are $E_{\text{OBC}}^{(j)} = 2J\cos\left({\pi j}/{(L+1})\right)$ and $E_{\text{PBC}}^{(j)}=2J\cos\left(2\pi j/L+i\bar{g}\right)$, respectively. For the PBC, the energy specrum is complex and forms an elliptical loop satisfying
 \begin{equation}
 \frac{(\text{Re}[E_{\text{PBC}}^{(j)}])^2}{\cosh^2\bar{g}}+\frac{(\text{Im}[ E_{\text{PBC}}^{(j)}])^2}{\sinh^2\bar{g}}=4J^2.
 \end{equation}
The spectral topology is characterized by the winding number of the loop of $E_{\text{PBC}}(k)=2J\cos\left(k+i\bar{g}\right)$ around the interior point $E_{\text{OBC}}^{(j)}$ \cite{Gong2018}: 
\begin{equation}
    \omega_j = \frac{1}{2\pi i} \oint_{0}^{2\pi} dk \, \partial_k \ln[E_{\text{PBC}}(k) - E_{\text{OBC}}^{(j)}].
\end{equation}
When $\bar{g}<0$ and $\bar{g}>0$, one has $\omega_j=\omega=\pm1$ as $E_{\text{PBC}}(k)$ traces clockwise and counterclockwise in the complex energy plane for all $j$, respectively; while $\omega=0$ for $\bar{g}=0$ as the PBC spectrum becomes real. The obtained topological phase diagram with respect to the winding number $\omega$ on the $W$-$h$ plane is shown in Fig. \ref{Efig1}(a).

The exact wavefunctions in Eq. (\ref{1Dwavefunction}) enable us to explore critical phases in thermodynamic limit $L\rightarrow\infty$. We use the fractal dimension $D_f$ related to the inverse
participation ratio (\text{IPR}) to quantify localization features \cite{Evers2008RMP}. For a wave function $\psi_n^{(j)}$, normalized as $\sum_{n=1}^{L} |\psi_n^{(j)}|^2 = 1$, its fractal dimension is defined as 
\begin{equation}
D_f^{(j)} = -\lim_{L \to \infty} [\ln \text{IPR}^{(j)} /\ln L], 
\end{equation}
with $\text{IPR}^{(j)} =\sum_{n=1}^{L} |\psi_n^{(j)}|^4$. In 1D systems, one has $D_f=1$ and $D_f=0$ for extended and localized states, respectively, whereas $0<D_f<1$ implies multifractal critical states. Since all eigenstates share the same feature here, three pure localization phases can be characterized by the average fractal dimension $D_f=\frac{1}{L} \sum_{j=1}^{L} D_f^{(j)}$. The localization phase diagram on the $W$-$h$ plane is obtained by computing $D_f$ under the PBC, as shown in Fig. \ref{Efig1}(b). One can find that the system is in the extended phase with $D_f \approx 1$ for $W<h$ and in the critical phase with $0<D_f<1$ for $W>h$. Note that we use the large-size scaling of the IPR to extract $D_f$ up to the lattice size $L>10^{9}$ with the exact wavefunctions $\psi_n^{(j)}$. According to Eq. (\ref{1Dwavefunction}), the average density distribution $\overline{\rho_n} = \frac{1}{L}\sum_{j=1}^L |\psi_n^{(j)}|^2\sim |\psi_n^{(j)}|^2$, such that all eigenstates of the system exhibit macroscopically multifractal profiles in the critical phase, as shown in Fig. \ref{Efig1}(c). In sharp contrast, in the CCP, the average density is almost spatially uniform with $\overline{\rho_n} \sim 1/L$ as the eigenstates exhibit distinct multifractal distributions, such as that at the critical point of the AA model shown in Fig. \ref{Efig1}(c).

We further study skin properties of the critical states. Under the OBC, as excepted, the critical eigenstates becomes boundary skin modes for $\omega\neq0$ when $\bar{g}\neq0$. Such a skin effect can be contributed to the fact that the cumulative imaginary phase of the wavefunctions $X_n$ (see Eq. (\ref{1Dwavefunction})) exhibits dominant increasing (decreasing) along the lattice when $\bar{g}<0$ ($\bar{g}>0$), as shown in Fig. \ref{Efig1}(d). In the globally reciprocal case of $\bar{g}=0$ with $\omega=0$, the multifractal critical eigenstates exhibit some main peaks in the lattice bulk for both OBC and PBC, which are located at some interfaces of $X_n$, as plotted in Fig. \ref{Efig1}(c). The interface-skin critical states also exhibit in case of $\bar{g}\neq0$ under the PBC, with dominant peaks being around the interfaces of $X_n-\bar{g}n$ (Fig. \ref{Efig1}(d)).

The physical origin of such a SCP with novel skin and localization properties is rooted in the extreme value statistics of random walks \cite{Schehr2012,Majumdar2020,Longhi2025} in the presence of quasiperiodic imaginary gauge fields $g_n$. For the case $\bar{g}=0$, the walk trajectory $X_n=\sum_{l=1}^{n-1} g_l$ fluctuates around zero with extreme points, where the sign of $g_n$ reverses and the interfaces of $X_n$ form. These interfaces induce a local skin effect on critical eigenstates, which globally maintain fractal characters imprinted from the quasiperiodic modulation of $X_n$. For the case $\bar{g} \neq 0$, only an extreme point of $X_n$ lies at the left or right end of the 1D lattice under the OBC, and thus boundary skin modes appear. Under the PBC, the periodic geometry eliminates the global trend of $X_n$ and allows the walk trajectory $X_n - n\bar{g}$ with quasiperiodic fluctuations [see Fig. \ref{Efig1}(d)], such that the local interface-skin feature of critical states persists regardless of $\bar{g}$ and for any $\omega$. The interfaces of $X_n - n\bar{g}$ determines the peaks of the wavefunctions in the SCP under the PBC.

Figure \ref{Efig1}(e) summarizes all characteristic properties of the SCP and CCP for comparisons. Unlike the CCP with overall uniform density distributions and the NHSE with eigenstate accumulation at open boundaries, the SCP is marked by a macroscopically multifractal distribution with all critical eigenstates sharing an identical profile and always accumulating at specific bulk interfaces under the PBC, which become topology-dependent boundary or interface skin modes under the OBC. The ballistic dynamics in the SCP furthermore distinguishes it from the CCP with diffusive dynamics.

\clearpage
\onecolumngrid
\setcounter{equation}{0}
\setcounter{figure}{0}
\setcounter{table}{0}
\setcounter{section}{0}
\renewcommand{\theequation}{S\arabic{equation}}
\renewcommand{\thefigure}{S\arabic{figure}}
\renewcommand{\thetable}{S\arabic{table}}
\renewcommand{\thesection}{S\arabic{section}}

\vspace{1em}
\begin{center}
\textbf{\large Supplemental Material}
\end{center}
\vspace{1em}

\subsection{A. Derivations of exact wavefunctions}

\textit{General case.---}In this section, we present a systematic and detailed analytical derivation to obtain the exact wavefunctions and energy spectrum of the $d$-dimensional non-Hermitian Hatano-Nelson model. We first investigate the general case where the imaginary gauge phase $g_\mathbf{r}$ is independent at each lattice site $\mathbf{r}$. In the general case, the $d$-dimensional non-Hermitian Hamiltonian is given by  
\begin{equation}
    \begin{split}
        \hat{H}  = \sum_{\alpha=1}^{d} \sum_{\mathbf{r}} J \Big[e^{-(X_{\mathbf{r}+\mathbf{e}_\alpha} - X_{\mathbf{r}})} \hat{c}_{\mathbf{r}}^{\dagger} \hat{c}_{\mathbf{r}+\mathbf{e}_\alpha} + e^{X_{\mathbf{r} +\mathbf{e}_\alpha} - X_{\mathbf{r}}} \hat{c}_{\mathbf{r}+\mathbf{e}_\alpha}^{\dagger} \hat{c}_{\mathbf{r}} \Big] + \hat{H}_B.
    \end{split}
    \label{eq1}
\end{equation}
Here $X_{\mathbf{r}} = \sum_{l_1=1}^{r_1} \cdots \sum_{l_d=1}^{r_d} g_{\mathbf{l}}$ is the d-dimensional cumulative imaginary gauge phase. $\hat{H}_B$ controls the boundary condition: $\hat{H}_B = 0$ for the OBC, whereas for the PBC,
\begin{equation}
\hat{H}_B  = \sum_{\alpha=1}^{d} \sum_{\{\mathbf{r} \mid r_\alpha = L\}} J [\exp(-(X_{\mathbf{r}+\mathbf{e}_\alpha} - X_{\mathbf{r}})) \hat{c}_{\mathbf{r}}^{\dagger} \hat{c}_{\mathbf{r} - (L-1)\mathbf{e}_\alpha} + \exp(X_{\mathbf{r} +\mathbf{e}_\alpha} - X_{\mathbf{r}}) \hat{c}_{\mathbf{r} - (L -1)\mathbf{e}_\alpha}^{\dagger} \hat{c}_{\mathbf{r}} ],
\label{eq2}
\end{equation}
where $\{ \mathbf{r} | r_\alpha = L \}$ denotes the set of all lattice sites at the boundary in the $\alpha$-th direction. The single-particle eigen-equation derived from the Hamiltonian (\ref{eq1}) is given by 
\begin{equation}
    \begin{split}
        E \psi_{\mathbf{r}}  = \sum_{\alpha=1}^{d}  J \Big[e^{X_{\mathbf{r}} - X_{\mathbf{r}-\mathbf{e}_\alpha}} \psi_{\mathbf{r}-\mathbf{e}_\alpha} + e^{-(X_{\mathbf{r} +\mathbf{e}_\alpha} - X_{\mathbf{r}})} \psi_{\mathbf{r}+\mathbf{e}_\alpha} \Big].
    \end{split}
    \label{eq3}
\end{equation}
For the OBC and PBC, the wavefunction satisfies $\psi_{\{\mathbf{r}|r_{\alpha}=0\}}=\psi_{\{\mathbf{r}|r_{\alpha}=L+1\}}=0$ and $\psi_\mathbf{r}=\psi_{\mathbf{r}+L\mathbf{e}_\alpha}$, respectively. To eliminate the non-reciprocity, we introduce an imaginary gauge transformation \cite{Midya2024,Longhi2025}
\begin{equation}
    \psi_{\mathbf{r}} = \phi_{\mathbf{r}} \exp(X_{\mathbf{r}}). 
\label{eq4}
\end{equation}
Thus, the original non-Hermitian spectral problem can be reduced to a Hermitian one:
\begin{equation}
E \phi_{\mathbf{r}} = \sum_{\alpha=1}^{d} J \left( \phi_{\mathbf{r}-\mathbf{e}_\alpha} + \phi_{\mathbf{r}+\mathbf{e}_\alpha} \right).
\label{eq5}
\end{equation}
For a $d$-dimensional eigen-equation, an exact solution is generally available only when the variables are separable. Whether separation of variables is possible depends on the boundary conditions. Under OBC, the wavefunction satisfies $\phi_{\{\mathbf{r}\mid r_{\alpha}=0\}} = \phi_{\{\mathbf{r}\mid r_{\alpha}=L+1\}} = 0$. Hence, the $d$-dimensional wavefunction can be expressed as a direct product of wavefunctions along each direction $\phi_{\mathbf{r}} = \prod_{\alpha=1}^{d} \phi^{(\alpha)}_{r_\alpha}$, while the corresponding eigenenergy is a simple summation $E = \sum_{\alpha=1}^{d} E^{(\alpha)}$, enabling variable separation and an analytical solution. Thus, we obtain independent 1D equations along each direction
\begin{equation}
E^{(\alpha)} \phi^{(\alpha)}_{r_\alpha} = J \left( \phi^{(\alpha)}_{r_\alpha - 1} + \phi^{(\alpha)}_{r_\alpha + 1} \right),
\label{eq6}
\end{equation}
with the boundary condition $\phi^{(\alpha)}_0 = \phi^{(\alpha)}_{L+1} = 0$. Consequently, the solution takes a standing-wave form $\phi_{r_\alpha}^{(\alpha)} = \sin(q^{(\alpha)} r_\alpha)$, which yields the eigenenergy $E^{(\alpha)} = 2J \cos q^{(\alpha)}$ with the wavevector $q^{(\alpha)}$. The boundary condition $\phi_{L+1}^{(\alpha)} = \sin(q^{(\alpha)} (L+1)) = 0$ quantizes the wavevector as $q^{(\alpha)} = \pi j^{(\alpha)}/(L+1)$, with $j^{(\alpha)} = 1, 2, \dots, L$. Thus, the exact solutions of the eigenenergies and eigenstates (up to a normalized constant) under OBC are given by \cite{Midya2024,Longhi2025}
\begin{equation}
\begin{cases}
    E^{(\alpha)}_j = 2J \cos(\pi j^{(\alpha)} /(L+1)), \\
    \phi_{r_\alpha}^{(j^{(\alpha)})} = \sin( \pi j^{(\alpha)}  r_\alpha/(L+1)).
\end{cases}
\label{eq7}
\end{equation}
Combining the solutions in all directions and applying the inverse gauge transformation, the exact solutions under OBC is given by 
\begin{equation}
\begin{cases}
E_{\mathrm{OBC}}^{\{j^{(\alpha)}\}} = \displaystyle\sum_{\alpha=1}^{d} 2J \cos\Bigl[\pi j^{(\alpha)} / (L+1)\Bigr], \\[10pt]
\psi_{\mathbf{r}}^{\{j^{(\alpha)}\}} = \displaystyle\prod_{\alpha=1}^{d} \sin\Bigl[\pi j^{(\alpha)} r_\alpha / (L+1)\Bigr] \, e^{X_{\mathbf{r}}}.
\end{cases}
\label{eq8}
\end{equation}
For the PBC, the original boundary condition $\psi_{\mathbf{r}} =\psi_{\mathbf{r}+L\mathbf{e}_\alpha} $ imposes a twisted boundary condition on the transformed wavefunction
\begin{equation}
   \phi_{\mathbf{r}+L\mathbf{e}_\alpha} = \phi_{\mathbf{r}} \exp(X_{\mathbf{r}}-X_{\mathbf{r}+L\mathbf{e}_\alpha}) = \phi_{\mathbf{r}} \exp(-X_{\{\mathbf{r} | r_\alpha = L \}}),  
   \label{eq9}
\end{equation} 
where $X_{\{\mathbf{r} | r_\alpha = L \}}$ denotes the cumulative imaginary gauge phases of all lattice sites at the boundary in the $\alpha$ direction. In general, it is a variable that depends on the remaining $d-1$ directions, which destroys the independence of variables along different directions. To apply the separation of variables method, the cumulative imaginary gauge phase at the
boundary must be identical for every direction, imposing the condition $X_{\{\mathbf{r} \mid r_\alpha = L\}}=X_{L}$. Under this condition, the boundary condition simplifies as $\phi_{\mathbf{r}+L\mathbf{e}_\alpha} = \phi_{\mathbf{r}} e^{-X_{L}}$, which is separable in each direction. In this case, the equation (\ref{eq6}) with the twisted boundary condition $\phi_{r_{\alpha}+L}^{(\alpha)}=\phi_{r_{\alpha}}^{(\alpha)}\exp(-X_L)$ can be solved exactly. Assuming a trial solution of the form $\phi_{r_\alpha}^{(\alpha)} = \exp(i q^{(\alpha)} r_\alpha - \theta^{(\alpha)} r_\alpha)$, the eigenenergy is obtained as $E^{(\alpha)} = 2J \cos(q^{(\alpha)} + i \theta^{(\alpha)})$. By substituting this ansatz into the boundary conditions, we find the relation $\exp(i q^{(\alpha)} L - \theta^{(\alpha)} L) = \exp(-X_{L}^{(\alpha)})$. Therefore, it follows that $q^{(\alpha)} = 2\pi j^{(\alpha)}/L$ and $\theta^{(\alpha)} = X_{L}^{(\alpha)}/L = \sum_{l=1}^{L} g_l^{(\alpha)}/L = \bar{g}^{(\alpha)}$, with $j^{(\alpha)} = 1, 2, \dots, L$, where $\bar{g}^{(\alpha)} = X_{L}^{(\alpha)}/L$ is the average imaginary gauge field in the $\alpha$-th direction. Thus, the exact solutions of the eigenenergies and eigenstates (up to a normalized constant) under the PBC are given by
\begin{equation}
\begin{cases}
    E^{(\alpha)}_j = 2J \cos( 2\pi j^{(\alpha)}/L + i \bar{g}^{(\alpha)} ), \\
    \phi_{r_\alpha}^{(j^{(\alpha)})} = \exp( i 2\pi j^{(\alpha)} r_\alpha/L - \bar{g}^{(\alpha)} r_\alpha ).
\end{cases}
\label{eq10}
\end{equation}
Consequently, the exact solution for the original Hamiltonian is given by
\begin{equation}
\begin{cases}
E_{PBC}^{\{j^{(\alpha)}\}} = \sum_{\alpha=1}^{d}2J \cos[ 2\pi j^{(\alpha)}/L + i \bar{g}^{(\alpha)} ], \\[10pt]
\psi_{\mathbf{r}}^{\{j^{(\alpha)}\}} = \prod_{\alpha=1}^{d} e^{ i 2\pi j^{(\alpha)} r_\alpha/L - \bar{g}^{(\alpha)} r_\alpha +X_{\mathbf{r}}}.
\end{cases}
\label{eq11}
\end{equation}

\textit{Degenerated case.---}When the imaginary gauge field along each direction $\alpha$ depends exclusively on its own directional coordinate $r_\alpha$, expressed as $g_{\mathbf{r}}^{(\alpha)} \mapsto g_{r_\alpha}^{(\alpha)}$, the variables in the gauge phase automatically decouple. This simplifies the total cumulative gauge phase into a sum of independent directional contributions: $X_{\mathbf{r}} \mapsto \sum_{\alpha=1}^{d} X_{r_{\alpha}}^{(\alpha)}$, where $X_{r_{\alpha}}^{(\alpha)} = \sum_{l=1}^{r_\alpha - 1} g_l^{(\alpha)}$. For this degenerated case, the Hamiltonian in Eq.~(\ref{eq1}) reduces to:
\begin{equation}
\hat{H} = \sum_{\alpha=1}^{d} \sum_{\mathbf{r}} J\left[ e^{-g_{r_\alpha}^{(\alpha)}} \hat{c}_{\mathbf{r}}^{\dagger} \hat{c}_{\mathbf{r} + \mathbf{e}_\alpha} +  e^{g_{r_\alpha}^{(\alpha)}} \hat{c}_{\mathbf{r} + \mathbf{e}_\alpha}^{\dagger} \hat{c}_{\mathbf{r}} \right] + \hat{H}_B,
\label{eq19}
\end{equation}
where the boundary term $\hat{H}_B = 0$ for OBC, and for PBC, it reads:
\begin{equation}
\hat{H}_B = \sum_{\alpha=1}^{d} \sum_{\{ \mathbf{r} \mid r_\alpha = L \}} J\left[  e^{-g_{L}^{(\alpha)}} \hat{c}_{\mathbf{r}}^{\dagger} \hat{c}_{\mathbf{r} - (L-1)\mathbf{e}_\alpha} +  e^{g_{L}^{(\alpha)}} \hat{c}_{\mathbf{r} - (L-1)\mathbf{e}_\alpha}^{\dagger} \hat{c}_{\mathbf{r}} \right].
\label{eq20}
\end{equation}
The single-particle eigen-equation for the Hamiltonian (\ref{eq19}) is given by 
\begin{equation}
E \psi_{\mathbf{r}} = \sum_{\alpha = 1}^{d} J \left( e^{g_{r_{\alpha} - 1}^{(\alpha)}} \psi_{\mathbf{r} - \mathbf{e}_{\alpha}} + e^{-g_{r_{\alpha}}^{(\alpha)}} \psi_{\mathbf{r} + \mathbf{e}_{\alpha}} \right).
\label{eq21}
\end{equation}
By utilizing the decoupled imaginary gauge transformation
\begin{equation}
\psi_{\mathbf{r}} = \phi_{\mathbf{r}} \exp\left( \sum_{\alpha=1}^{d} X_{r_{\alpha}}^{(\alpha)} \right),
\label{eq22}
\end{equation}
the problem is mapped to the same Hermitian equation as Eq.~(\ref{eq5}). Since the gauge fields depend only on their respective directional coordinates, the separability of the wavefunctions is satisfied for both the OBC and PBC without requiring additional constraints on the cumulative imaginary gauge phase at the boundary. Following the same separation of variables procedure for the Hermitian counterpart, the transformed eigenstates $\phi_{\mathbf{r}}$ and eigenenergies retain the same form as the general case, but the original wavefunction $\psi_{\mathbf{r}}$ differs by a phase factor in the imaginary gauge transformation. The exact wavefunctions for the $d$-dimensional degenerated Hatano-Nelson model are given by
\begin{equation}
\psi_{\mathbf{r}}^{\{j^{(\alpha)}\}}=
\begin{cases}
\displaystyle\prod_{\alpha=1}^{d} \sin\Bigl[\frac{\pi j^{(\alpha)} r_\alpha}{L+1}\Bigr] \exp\Bigl(X_{r_\alpha}^{(\alpha)}\Bigr), & \text{OBC}; \\[4ex]
\displaystyle\prod_{\alpha=1}^{d} \exp\Bigl( i \frac{2\pi j^{(\alpha)} r_\alpha}{L} - \bar{g}^{(\alpha)} r_\alpha +X_{r_\alpha}^{(\alpha)} \Bigr), & \text{PBC},
\end{cases}
\label{eq23}
\end{equation}
where $j^{(\alpha)} = 1, 2, \dots, L$. The corresponding eigenenergies take the form $E_{\mathrm{OBC}}^{\{j^{(\alpha)}\}} = \sum_{\alpha=1}^{d}2J \cos[\pi j^{(\alpha)} /(L+1)]$ and $E_{\mathrm{PBC}}^{\{j^{(\alpha)}\}} = \sum_{\alpha=1}^{d}2J \cos[ 2\pi j^{(\alpha)}/L + i \bar{g}^{(\alpha)} ]$, respectively.

For the simplest 1D case, we set $d = \alpha = 1$ and denote $g_\textbf{r}^{(1)} = g_n$ with the lattice site $n \in [1, L]$. The Hamiltonian in Eq.~(\ref{eq19}) becomes
\begin{equation}
\hat{H}_{1D} = J\sum_{n=1}^{L-1} \left( e^{-g_n} \hat{c}_{n}^\dagger \hat{c}_{n+1} + e^{g_n} \hat{c}_{n+1}^\dagger \hat{c}_{n} \right) + \hat{H}_B,
\label{eq24}
\end{equation}
where $\hat{H}_B = 0$ for the OBC and $\hat{H}_B = J(e^{-g_L} \hat{c}_L^\dagger \hat{c}_1 + e^{g_L} \hat{c}_1^\dagger \hat{c}_L)$ for the PBC. In this case, the imaginary gauge transformation in Eq.~(\ref{eq22}) reduces to $\psi_n = \phi_n e^{X_n}$ with $X_n = \sum_{l=1}^{n-1} g_l$. The solutions are obtained directly from the general results in Eqs.~(\ref{eq23}) by setting $r_\alpha \to n$, $j^{(\alpha)} \to j$, and $\bar{g}^{(\alpha)} \to \bar{g} \equiv L^{-1}\sum_{l=1}^L g_l$. This yields the exact wavefunctions in the 1D generalized Hatano-Nelson model with site-dependent imaginary gauge phase:
\begin{equation}
\psi_{n}^{(j)}=
\begin{cases}
    \sin\left( \dfrac{\pi j n}{L+1} \right) e^{X_{n}}, & \text{OBC}; \\[2ex]
    \exp\left( i \dfrac{2\pi j n}{L} - \bar{g} n + X_{n} \right), & \text{PBC},
\end{cases}
\label{eq25}
\end{equation}
up to a normalized constant. The corresponding eigenenergies are $E_{\text{OBC}}^{(j)} = 2J \cos[\pi j /(L+1)]$ and $E_{\text{PBC}}^{(j)} = 2J \cos[ 2\pi j/L + i \bar{g} ]$, for $j=1,2,\dots,L$.

\subsection{B. Construction of imaginary gauge phase pattern}

For clarity, we focus on the 2D case as illustrated in Fig. 1 of the main text. We set $d = 2$ and denote $X_{\mathbf{r}}=S_{n,m}$ and $g_{\mathbf{r}} = g_{n,m}$ with the lattice site $n,m \in [1, L]$ in the two respective directions. The model Hamiltonian in Eq.~(\ref{eq1}) is then reduced to
\begin{equation}
    \begin{split}
        \hat{H}_{2D}  = \sum_{n,m=1}^{L-1} J [ e^{-(S_{n+1,m} - S_{n,m})} \hat{c}_{n,m}^{\dagger} \hat{c}_{n+1,m} + e^{S_{n+1,m} - S_{n,m} } \hat{c}_{n+1,m}^{\dagger} \hat{c}_{n,m} \\
        +e^{-(S_{n,m+1} - S_{n,m})} \hat{c}_{n,m}^{\dagger} \hat{c}_{n,m+1} + e^{S_{n,m+1} - S_{n,m} } \hat{c}_{n,m+1}^{\dagger} \hat{c}_{n,m}] + \hat{H}_B,
    \end{split}
    \label{eq12}
\end{equation}
where $S_{n,m} = \sum_{l_1=1}^{n} \sum_{l_2=1}^{m} g_{l_1,l_2}$. For the PBC, $\hat{H}_B  = \sum_{n,m=1}^{L} J [e^{-(S_{L+1,m} - S_{L,m})} \hat{c}_{L,m}^{\dagger} \hat{c}_{1,m} + e^{S_{L+1,m} - S_{L,m}} \hat{c}_{1,m}^{\dagger} \hat{c}_{L,m}
+e^{-(S_{n,L+1} - S_{n,L})} \hat{c}_{n,L}^{\dagger} \hat{c}_{n,1} + e^{S_{n,L+1} - S_{n,L}} \hat{c}_{n,1}^{\dagger} \hat{c}_{n,L} ]$. For the OBC, $\hat{H}_B = 0$. The single-particle eigen-equation in this 2D case is written as
\begin{equation}
    \begin{split}
        E \psi_{n,m}  =   J \Big[e^{S_{n,m} - S_{n-1,m}} \psi_{n-1,m} + e^{-(S_{n+1,m} - S_{n,m})} \psi_{n+1,m} \\
        +e^{S_{n,m} - S_{n,m-1}} \psi_{n,m-1} + e^{-(S_{n,m+1} - S_{n,m})} \psi_{n,m+1} \Big].
    \end{split}
    \label{eq13}
\end{equation}
For OBC, the wavefunction satisfies $\psi_{0,m}=\psi_{L+1,m}=\psi_{n,0}=\psi_{n,L+1}=0$, while for PBC, it satisfies $\psi_{n,m}=\psi_{n+L,m}=\psi_{n,m+L}$, respectively. To gauge away the non-reciprocity, we introduce an imaginary gauge transformation
\begin{equation}
   \psi_{n,m} = \phi_{n,m} \exp(S_{n,m}).
    \label{eq14}
\end{equation}
The original non-Hermitian spectral problem can then be mapped to a Hermitian one
\begin{equation}
    \begin{split}
        E \phi_{n,m}  =   J \Big[\phi_{n-1,m} +  \phi_{n+1,m} 
        + \phi_{n,m-1} +  \phi_{n,m+1} \Big].
    \end{split}
    \label{eq15}
\end{equation}
Under the OBC, the wavefunction satisfies $\phi_{0,m}=\phi_{L+1,m}=\phi_{n,0}=\phi_{n,L+1}=0$. Hence, the 2D wavefunction can be expressed as a direct product of wavefunctions along each direction $\phi_{n,m} = \phi_n \phi_m$, which enables variable separation. The exact solutions for the OBC is given by 
\begin{equation}
\begin{cases}
E_{\mathrm{OBC}}^{(j^{(1)},j^{(2)})} =  2J [\cos(\pi j^{(1)} / (L+1))+\cos(\pi j^{(2)} / (L+1))], \\[10pt]
\psi_{\mathbf{r}}^{(j^{(1)},j^{(2)})} =  \sin(\pi j^{(1)} n / (L+1))\sin(\pi j^{(2)} m / (L+1))e^{S_{n,m}}.
\end{cases}
\label{eq16}
\end{equation}
For the PBC, the original boundary condition $\psi_{n,m}=\psi_{n+L,m}=\psi_{n,m+L}$ imposes a twisted boundary condition on the transformed wavefunction
\begin{equation}
    \begin{cases}
   \phi_{n+L,m} = \phi_{n,m} \exp(S_{n,m}-S_{n+L,m}) = \phi_{n,m} \exp(-S_{L,m}), \\
   \phi_{n,m+L} = \phi_{n,m} \exp(S_{n,m}-S_{n,m+L}) = \phi_{n,m} \exp(-S_{n,L}).
   \end{cases}
   \label{eq17}
\end{equation} 
Thus, we obtain the relationship $S_{L,m}=S_{n,L}\equiv \tilde{S}$, which indicates that all the cumulative imaginary gauge phases at the boundary are identical for every direction, denoted as $\tilde{S}$. The boundary condition simplifies to $\phi_{n+L,m} = \phi_{n,m} e^{-\tilde{S}}$ and $\phi_{n,m+L} = \phi_{n,m} e^{-\tilde{S}}$, which is separable in each direction and gives rise to the exact solutions:
\begin{equation}
\begin{cases}
E_{PBC}^{(j^{(1)},j^{(2)})} = 2J [\cos(2\pi j^{(1)}/L + i \bar{g}^{(1)} )+\cos(2\pi j^{(2)}/L + i \bar{g}^{(2)} )], \\[10pt]
\psi_{n,m}^{(j^{(1)},j^{(2)})} =  e^{ i 2\pi j^{(1)} n/L+i 2\pi j^{(2)} m/L - n\tilde{S}/L- m\tilde{S}/L  +S_{n,m}}.
\end{cases}
\label{eq18}
\end{equation}
Here we have used $\bar{g}^{(1)} =\bar{g}^{(2)} = \tilde{S}/L$ for the average imaginary gauge phase in each direction.

\begin{figure}[t]
    \centering
     \includegraphics[width=0.9\linewidth]{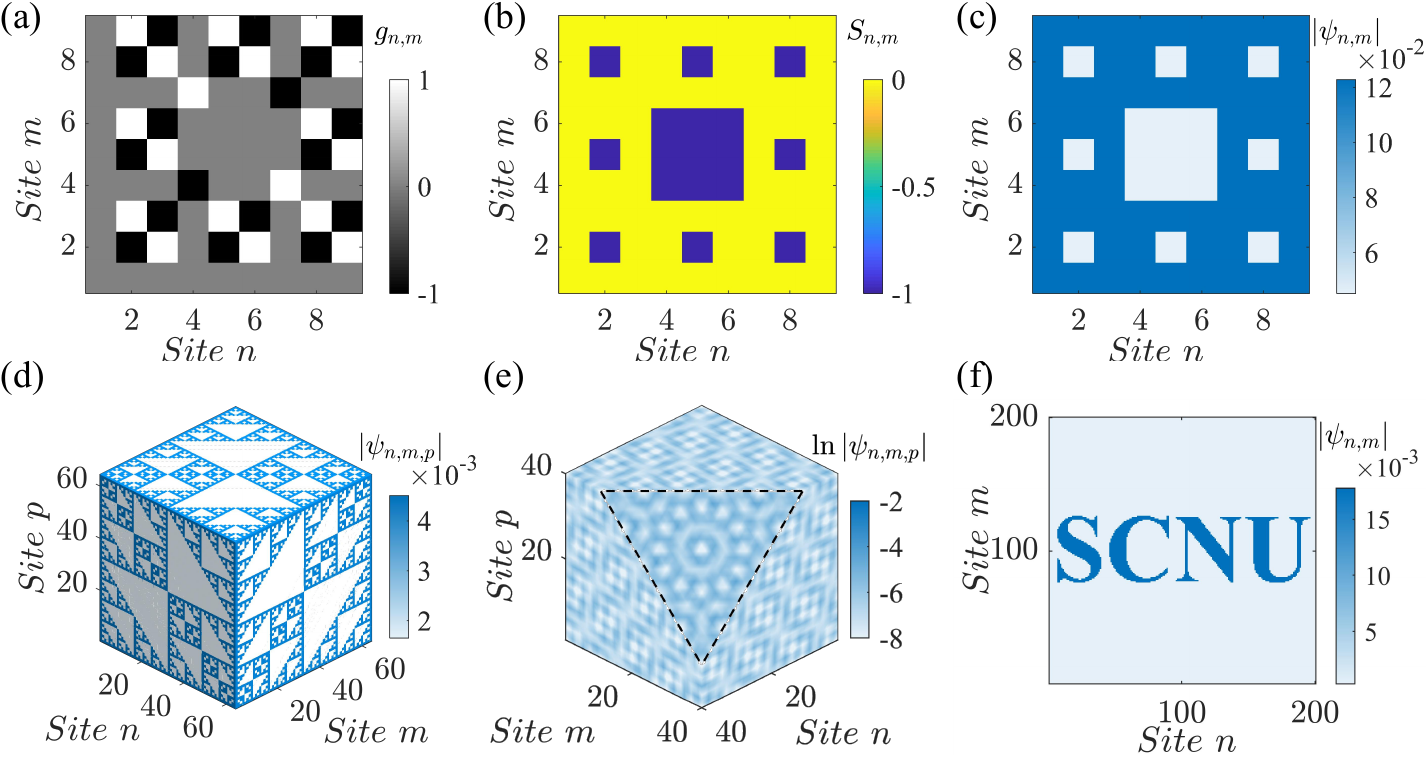}
     \caption{Configurable wavefunctions. Profiles of the 2D imaginary gauge phase $g_{n,m}$ (a), cumulative imprint phase $S_{n,m}$ (b), and imprinted wavefunction profile $|\psi_{n,m}|$ (c) with the Sierpiński-carpet structure. Exact wavefunctions engineered into a 3D glued Sierpiński triangle structure (d), a 3D incommensurate moir\'e pattern (e), and the ``SCNU'' letters (f).}
     \label{figS1}
\end{figure}

Based on the exact solutions and the IGPI framework (see Fig. 1 in the main text), we can engineer arbitrarily configurable wavefunctions. Taking the 2D case as an example, the spatial profile of the wavefunction is primarily governed by the cumulative phase $S_{n,m}$ under both boundary conditions. This allows us to arbitrarily design the distribution of $S_{n,m}$ to imprint the desired pattern into the 2D exact wavefunctions. It is noteworthy that under the PBC, the term $e^{-(n+m)\tilde{S}/L}$ of the wavefunction in Eq. (\ref{eq18}) may modify the distribution profile exponentially. To ensure that the wavefunction profile is determined solely by $S_{n,m}$, we typically constrain the cumulative phase at the boundaries to vanish by imposing $\tilde{S}=0$. Taking the 2D Sierpiński-carpet fractal as an example, we first design the distribution of $S_{n,m}$ to match the fractal pattern while maintaining zero boundary values, as shown in Fig.~\ref{figS1}(b). The imaginary gauge field $g_{n,m}$ is then obtained via the discrete difference relation 
\begin{equation}
g_{n,m}=S_{n,m}-S_{n-1,m}-S_{n,m-1}+S_{n-1,m-1}, 
\end{equation}
with its distribution as displayed in Fig.~\ref{figS1}(a). The resulting Sierpiński-carpet fractal pattern of wavefunctions is shown in Fig.~\ref{figS1}(c).

The same imaginary phase imprint procedure can be implemented in the 3D case with the imaginary gauge phase 
\begin{equation}
g_{n,m,p}=V_{n,m,p}-V_{n-1,m,p}-V_{n,m-1,p}-V_{n,m,p-1}+V_{n-1,m-1,p}+V_{n-1,m,p-1}+V_{n,m-1,p-1}-V_{n-1,m-1,p-1}. 
\end{equation}
To ensure the existence of an exact solution, the 3D cumulative gauge field is assumed to be uniform on the boundaries with $V_{L,m,p} = V_{n,L,p} = V_{n,m,L} = 0$. For clarity, this uniform outermost boundary layer is omitted from the visualization to prevent obstruction and better showcase the wavefunction profile. By constructing the corresponding imaginary gauge phase pattern, we obtain fractal wavefunctions with a 3D glued Sierpiński triangle structure, as illustrated in Fig.~\ref{figS1}(d). This structure is formed by concatenating four 3D Sierpiński triangles, demonstrating that complex fractal states can be engineered in non-fractal lattices. Additionally, we design moir\'e wavefunctions in a regular lattice as shown in Fig.~4 of the main text. Both 2D and 3D moiré potentials are formed by the superposition of original lattice potentials subjected to spatial rotations. In 2D systems, the primitive square lattice potential is defined as $V(\mathbf{r}) = \cos(\pi x) + \cos(\pi y)$ (see Region I in Fig. 4(b) in the main text) \cite{PWang2019}. Introducing a 2D rotation matrix $\mathbf{S}(\theta)$ ($\theta$ represents the rotation angle), the moiré potential is expressed as $V_{\text{2-layers}}(\mathbf{r}) = \left[ V(\mathbf{r}) + V(\mathbf{S}(\theta)\mathbf{r}) \right]^2$. When the twist angle $\theta$ is a Pythagorean angle, the lattice satisfies the commensurate condition to form a periodic moiré potential. Conversely, it becomes an incommensurate moiré quasicrystal potential. If a third square lattice is introduced, the total superimposed potential of the trilayer lattice is given by $V_{\text{3-layers}}(\mathbf{r}) = \left[ V(\mathbf{r}) + V(\mathbf{S}(\theta_2)\mathbf{r}) + V(\mathbf{S}(\theta_3)\mathbf{r}) \right]^2$. When the relative twist angles between adjacent layers both satisfy the commensurate condition, they interfere to form a super-moiré potential \cite{Ren2025,Wang2019}. Further extending this 3D systems, the primitive cubic lattice potential is defined as $V = \sin^2(\pi x) + \sin^2(\pi y) + \sin^2(\pi z)$ \cite{Wang2024,Wang2025}. Introducing a 3D rotation matrix $\mathbf{R}(\mathbf{L}, \theta)$, where $\mathbf{L}$ denotes the spatial rotation axis, the resulting 3D moiré potential generated by the interference of two cubic lattice potentials is written as $V_{\text{3D}}(\mathbf{r}) = V(\mathbf{r}) + V(\mathbf{R}(\mathbf{L}, \theta)\mathbf{r})$. Under the commensurate condition (e.g., rotating by $\theta = \pi/3$ about the body diagonal axis $\mathbf{L} = [1,1,1]$), the system forms a 3D moiré periodic potential. Otherwise, it forms a 3D incommensurate moiré potential as shown in Fig. ~\ref{figS1}(e) ($\theta=\pi/6$) \cite{Wang2024,Wang2025}.

Furthermore, we customize the wavefunction to spell out the letters ``SCNU", as depicted in Fig.~\ref{figS1}(f). These examples demonstrate that the IGPI framework enables us to engineer wavefunctions of arbitrary profiles in simple regular lattices, extending far beyond fractal and moir\'e structures.

\subsection{C. Topological phase boundaries and critical exponents}

In this section, we analytically derive the topological phase boundaries shown in Fig. 2(a) of the main text for the 1D Hamiltonian (\ref{eq24}), and numerically extract the critical exponents from finite-size scaling. At the topological transition point, the localization length $\Lambda$ under the OBC diverges due to the gap-closing nature, which corresponds to $\Lambda^{-1} \rightarrow 0$. The inverse of localization length is expressed as
\begin{equation}
    \begin{aligned}
        \Lambda^{-1} &= \lim_{L\rightarrow\infty}\frac{1}{L}\ln\left|\frac{\psi_{L}}{\psi_{1}}\right| 
        = \lim_{L\rightarrow\infty}\frac{1}{L}\ln\left|\frac{\sin\left(\frac{\pi j L}{L+1}\right) e^{X_{L}}}{\sin\left(\frac{\pi j}{L+1}\right) e^{X_{1}}}\right| 
        = \lim_{L\rightarrow\infty}\frac{|X_{L}|}{L} \\ 
        &\approx \lim_{L\rightarrow\infty}\frac{1}{L}\sum_{n=1}^{L} g_{n} 
        = \lim_{L\rightarrow\infty}\frac{1}{L}\sum_{n=1}^{L}\ln\left|W\cos(2\pi \beta n) + h\cos(\pi  n)\right| 
        = \bar{g},
    \end{aligned}
\label{eq26}
\end{equation}
where $\beta= (\sqrt{5}-1)/2$ is an irrational number. Thus, the topological transition occurs at critical points with $\bar{g}=0$, lying between two topologically distinct phases with winding number $\omega=1$ for $\bar{g}<0$ and $\omega=-1$ for $\bar{g}>0$, which are consistent with the results in the main text.

\begin{figure}[t]
    \centering
    \includegraphics[width=0.9\linewidth]{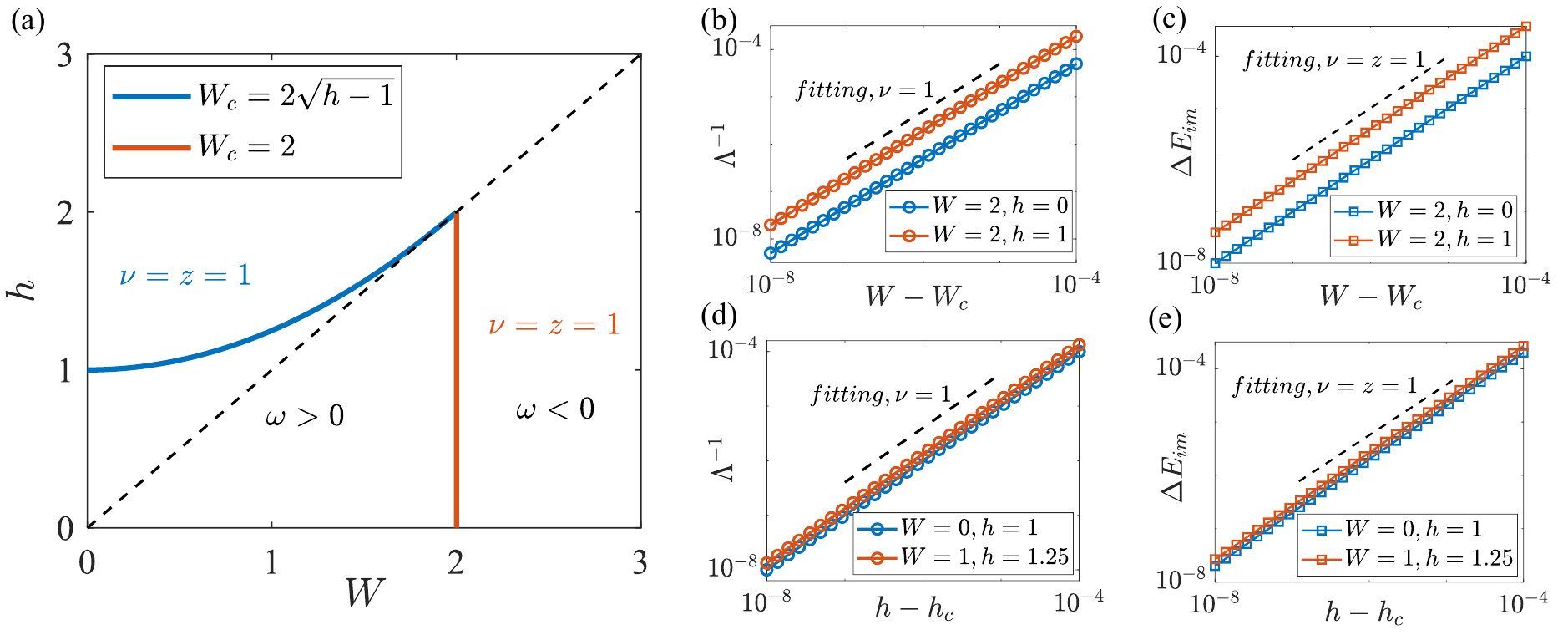}
    \caption{ (a) Exact topological boundaries from Eq. (\ref{eq29}) denoted by blue and red lines, which are separated by the dashed line $W = h$ and have critical exponents $\nu = z = 1$. (b,c) Scaling of $\Lambda^{-1}$ and $\Delta E_{im}$ for the red boundary at $(W=2,h=0)$ and $(W=2,h=1)$. (d,e) Scaling for the blue boundary at $(W=0,h=1)$ and $(W=1.25,h=1)$. The lattice size is $L=987$ for the finite-size scaling in (b-e).}
    \label{figS2}
\end{figure}

To further obtain the topological phase boundaries in the $W$-$h$ parameter plane, we utilize the irrational nature of $\beta$, which converts the discrete summation into a continuous integration:
\begin{equation}
    \begin{aligned}
        \Lambda^{-1} &= \lim_{L \rightarrow \infty} \frac{1}{L} \sum_{n=1}^{L} \ln \left|W \cos (2 \pi \beta n) + (-1)^{n} h\right| \\
        &= \lim_{L \rightarrow \infty} \frac{1}{L} \left[ \sum_{n \text { even }} \ln \left|W \cos (2 \pi \beta n) + h\right| 
        + \sum_{n \text { odd }} \ln \left|W \cos (2 \pi \beta n) - h\right| \right] \\
        &= \frac{1}{2} \int_{0}^{1} \ln \left|W \cos (2 \pi t) + h\right|  dt 
        + \frac{1}{2} \int_{0}^{1} \ln \left|W \cos (2 \pi t) - h\right|  dt \\
        &= \frac{1}{2} \cdot \frac{1}{2 \pi} \int_{0}^{2 \pi} \ln \left|W \cos x + h\right|  dx 
        + \frac{1}{2} \cdot \frac{1}{2 \pi} \int_{0}^{2 \pi} \ln \left|W \cos x - h\right|  dx \\
        &= \frac{1}{2 \pi} \int_{0}^{2 \pi} \ln \left|W \cos x + h\right|  dx,
    \end{aligned}
    \label{eq27}
\end{equation}
where the equality of the two integrals follows from the identity $\ln |W \cos (x+\pi) + h| =  \ln |W \cos x - h|$. This integral can be solved as
\begin{equation}
    \Lambda^{-1} =
    \begin{cases}
    \ln \left( \dfrac{|h| + \sqrt{h^{2} - W^{2}}}{2} \right), & |W| < |h|\\
    \ln \left( \dfrac{|W|}{2} \right). & |W| \geq |h|
    \end{cases}
    \label{eq28}
\end{equation}
The topological phase transition occurs when $\Lambda^{-1} = 0$, which gives the critical points:
\begin{equation}
    \begin{cases}
    |W_c| = 2\sqrt{h-1}, & |W| < |h| \\
    |W_c| = 2, & |W| \geq |h|
    \end{cases}
    \label{eq29}
\end{equation}
These analytical results in Fig. \ref{figS2}(a) are in exact agreement with the phase boundaries depicted in Fig. 2(a) of the main text.

We also investigate the critical behaviour near the topological transition by calculating the correlation-length critical exponent $\nu$ and the dynamical critical exponent $z$. In the large $L$ limit, both the inverse of localization length $\Lambda^{-1}$ and the imaginary energy gap $\Delta E_{im}$ exhibit the following scaling behaviour near the critical point
\begin{equation}
    \begin{cases}
        \Lambda^{-1} \sim |h - h_c|^\nu, \\ 
        \Delta E_{im} \sim |h - h_c|^{\nu z},
    \end{cases}
    \quad  |W| < |h|
    \label{eq30}
\end{equation}
and
\begin{equation}    
    \begin{cases}
        \Lambda^{-1} \sim |W - W_c|^\nu,  \\
        \Delta E_{im} \sim |W - W_c|^{\nu z}.
    \end{cases}
    \quad  |W| \geq |h|
    \label{eq31}
\end{equation}
We numerically calculate  $\Lambda^{-1}$ and $\Delta E_{im}$ near the critical points and extract the critical exponents through finite-size scaling analysis. As established in the main text, the eigenenergy is complex and forms an ellipse described by
\begin{equation}  
\frac{(\text{Re}[E_{\text{PBC}}^{(j)}])^2}{\cosh^2\bar{g}}+\frac{(\text{Im}[E_{\text{PBC}}^{(j)}])^2}{\sinh^2\bar{g}}=4J^2
\end{equation}   
for PBC. The corresponding imaginary point gap, which corresponds to the minor axis of the ellipse and persists for $\bar{g} \neq 0$, is given by $\Delta E_{\text{im}}=2J|\sinh\bar{g}|$. Numerical results along the critical lines are shown in Fig. \ref{figS2}(b,c) for $|W_c| = 2$ at $(W=2,h=0)$ and $(W=2,h=1)$, and in Figs. \ref{figS2}(d-e) for $|W_c| = 2\sqrt{h-1}$ at $(W=0,h=1)$ and $(W=1.25,h=1)$. In all cases, we obtain the same critical exponents $\nu = z = 1$, indicating that the topological transitions along both boundaries belong to the same universality class. This universality arises because both the inverse of localization length $\Lambda^{-1} \approx \bar{g}$ and the imaginary energy gap are governed by the average imaginary gauge field strength $\bar{g}$.

\subsection{D. Exact skin critical states in higher dimensions}

\begin{figure}[t]
    \centering
    \includegraphics[width=0.9\linewidth]{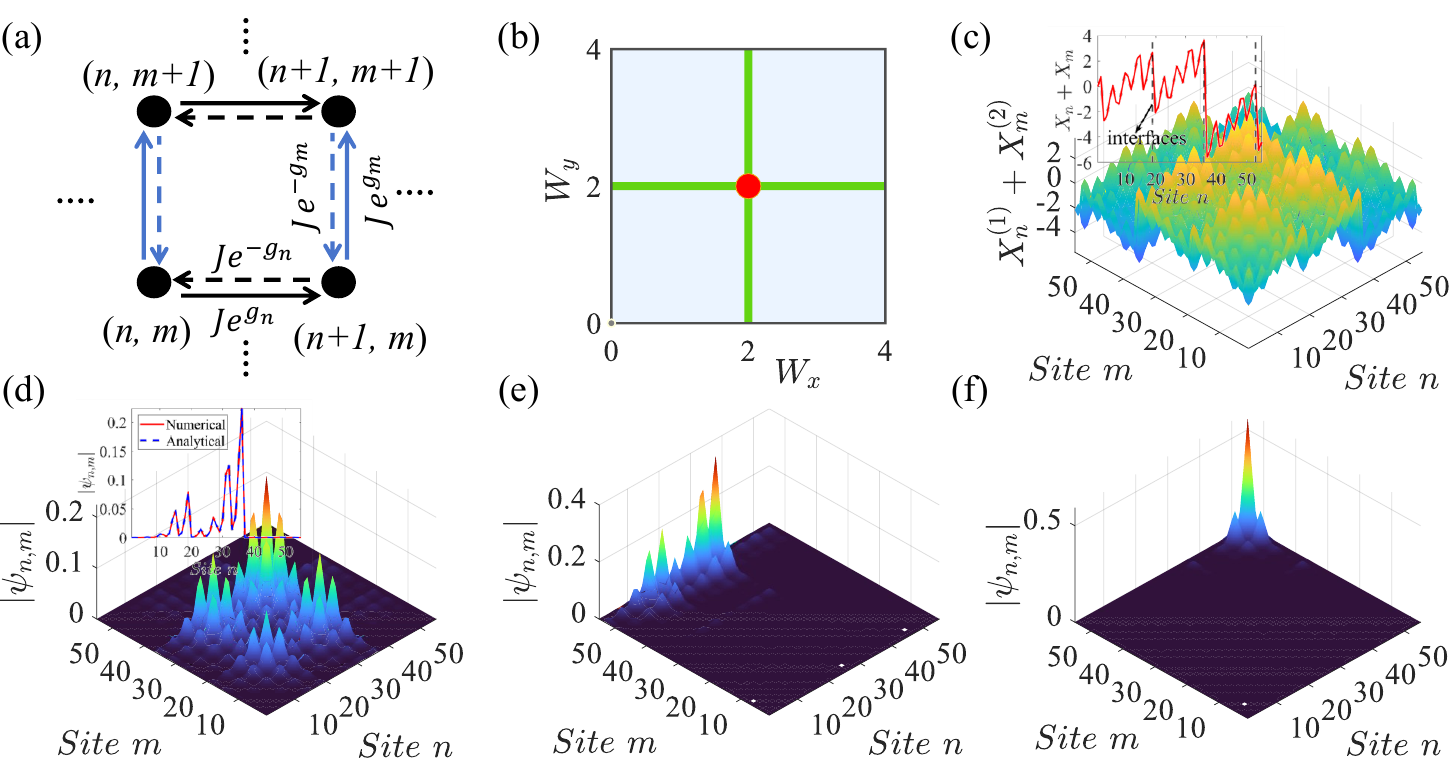}
    \caption{(a) Schematic of the 2D Hatano-Nelson model with a quasiperiodic imaginary gauge field. (b) Phase diagram of 2D skin critical states under the OBC. The red point, green lines, and blue regions denote the bulk-interface, line-boundary, and corner skin critical states, respectively. (c) Spatial distribution of the total imaginary phase $X_n + X_m$ for $W_x = W_y = 2$. Inset: Diagonal elements of total imaginary phase with interfaces marked by dashed lines. (d) Spatial distribution of 2D skin critical state. Inset: Analytical versus numerical wavefunctions distribution along the diagonal. (e) Spatial distribution of line-boundary skin critical state. (f) Representative corner skin mode localized at the upper-right lattice site.}
    \label{figS3}
\end{figure}

In this section, we extend the exact skin critical states to the higher dimensions. Here we focus on the 2D case for clarity, and note that our results can be directly extended to 3D lattices via the IGPI method with decoupled imaginary gauge phases along each directions [see Eqs. (\ref{eq22}) and (\ref{eq23})]. For the 2D case of $d = 2$, we denote $g_{r_1}^{(1)} = g_n^{(1)}$ and $g_{r_2}^{(2)} = g_m^{(2)}$ with the indices of the lattice site $n,m \in [1, L]$ in the two respective directions. As schematically shown in Fig. \ref{figS3}(a), the model Hamiltonian in Eq.~(\ref{eq19}) then reduces to the 2D counterpart:
\begin{equation}
\hat{H}_{2D} = J \sum_{n,m=1}^{L-1} \left( e^{-g_n^{(1)}} \hat{c}_{n,m}^\dagger \hat{c}_{n+1,m} + e^{g_n^{(1)}} \hat{c}_{n+1,m}^\dagger \hat{c}_{n,m} + e^{-g_m^{(2)}} \hat{c}_{n,m}^\dagger \hat{c}_{n,m+1} + e^{g_m^{(2)}} \hat{c}_{n,m+1}^\dagger \hat{c}_{n,m} \right) + \hat{H}_B,
\label{eq32}
\end{equation}
where the boundary term is $\hat{H}_B = 0$ for the OBC and $\hat{H}_B = \sum_{n,m=1}^{L} J(e^{-g_L^{(1)}} \hat{c}_{L,m}^\dagger \hat{c}_{1,m} + e^{g_{L}^{(1)}} \hat{c}_{1,m}^\dagger \hat{c}_{L,m}+e^{-g_L^{(2)}} \hat{c}_{n,L}^\dagger \hat{c}_{n,1} + e^{g_L^{(2)}} \hat{c}_{n,1}^\dagger \hat{c}_{n,L})$ for the PBC. The imaginary gauge transformation in Eq.~(\ref{eq22}) takes the 2D form $\psi_{n,m} = \phi_{n,m} \, e^{X_n^{(1)} + X_m^{(2)}}$, where $X_n^{(1)} = \sum_{l=1}^{n-1} g_l^{(1)}$ and $X_m^{(2)} = \sum_{l=1}^{m-1} g_l^{(2)}$. Following the same analysis as in the general $d$-dimensional case, the solutions follow directly from Eqs.~(\ref{eq23}), with the identifications $(r_\alpha, j^{(\alpha)}, \bar{g}^{(\alpha)}) \to (n, j^{(1)}, \bar{g}^{(1)})$ and $(m, j^{(2)}, \bar{g}^{(2)})$. Applying the inverse transformation yields the exact wavefunctions in the 2D generalized Hatano-Nelson model with site-dependent imaginary gauge phase:
\begin{equation}
\psi_{n,m}^{(j^{(1)},j^{(2)})}=
\begin{cases}
    \sin\left( \dfrac{\pi j^{(1)} n}{L+1} \right) \sin\left( \dfrac{\pi j^{(2)} n}{L+1} \right) e^{X_n^{(1)} + X_m^{(2)}}, & \text{OBC}; \\[2ex]
    \exp\left(i 2\pi j^{(1)} n/L + i 2\pi j^{(2)} m/L - \bar{g}^{(1)} n - \bar{g}^{(2)} m + X_n^{(1)} + X_m^{(2)} \right), & \text{PBC},
\end{cases}
\label{eq33}
\end{equation}
where $j^{(1)},j^{(2)}=1,2,\dots,L$. The corresponding eigenenergies are $E_{\text{OBC}}^{(j^{(1)},j^{(2)})} = 2J[\cos(\pi j^{(1)}/(L+1)) + \cos(\pi j^{(2)}/(L+1))]$ and $E_{\text{PBC}}^{(j^{(1)},j^{(2)})} = 2J[\cos(2\pi j^{(1)}/L + i\bar{g}^{(1)}) + \cos(2\pi j^{(2)}/L + i\bar{g}^{(2)})]$.

We consider the quasiperiodic imaginary gauge fields $g_n^{(1)} = \ln(|W_x \cos(2 \pi \beta n )|)$ and $ g_m^{(2)} = \ln(|W_y \cos(2 \pi \beta m )|)$, with $\beta = (\sqrt5-1)/2$ and $W_{x,y}$ being the quasiperiodic strengths along $x$ and $y$ directions. Under the PBC, all eigenstates in the $W_x$-$W_y$ parameter plane are characterized as 2D skin critical states accumulating at specific bulk interfaces, similar to the 1D case discussed in the main text. Under the OBC, the system exhibits a richer phase diagram as shown in Fig. \ref{figS3}(b). At $W_x = W_y = 2$, marked by the red point in the phase diagram, despite local non-reciprocity, the system remains globally reciprocal with $\bar{g}^{(1)}=\bar{g}^{(2)}=0$, and all eigenstates retain the characteristic of 2D bulk-interface skin critical states. For parameters lying on the green lines, where one direction is globally reciprocal and the other is non-reciprocal, the eigenstates display anisotropic profiles: Along the globally reciprocal direction the skin-fractal critical behaviour is preserved, while along the non-reciprocal direction skin localization occurs, namely the line-boundary skin critical states. In the remaining (blue) regions of the phase diagram in Fig. \ref{figS3}(b), where the hoppings along both two directions are globally non-reciprocal, the corner skin states emerge. The four separate regions correspond to corner states localized at the four distinct corners of the 2D lattice. The spatial profile of the eigenstates is governed primarily by the total imaginary phase accumulated by the hopping along the two directions $X_n^{(1)} + X_m^{(2)}$, as derived in the analytical solution Eq. (\ref{eq33}). The spatial distribution of total imaginary phase for the parameters $W_x = W_y = 2$ is plotted in Fig. \ref{figS3}(c), revealing a pronounced fractal structure of the imprinted imaginary gauge phase. The corresponding distribution of the diagonal elements of the total imaginary phase for $m=n$ is shown in the inset of Fig. \ref{figS3}(c), which further highlights the bulk interfaces marked by black dashed lines. The typical eigenstate distribution with the fractal pattern skinned to the bulk interfaces is shown in Fig. \ref{figS3}(d). Perfect coincidence of the analytical and numerical wave function distributions along the diagonal, as demonstrated in the inset of Fig. \ref{figS3}(d), confirming the validity of the exact solutions in this 2D system. A line-boundary skin critical mode is also presented in Fig. \ref{figS3}(e), where the wavefunction exhibits  critical property along the $x$-direction and skin localization along the $y$-direction. A representative corner skin mode is shown in Fig. \ref{figS3}(f), with the wavefunction sharply localized at the upper-right corner of the 2D system.

\begin{figure}[t]
    \centering
     \includegraphics[width=0.95\linewidth]{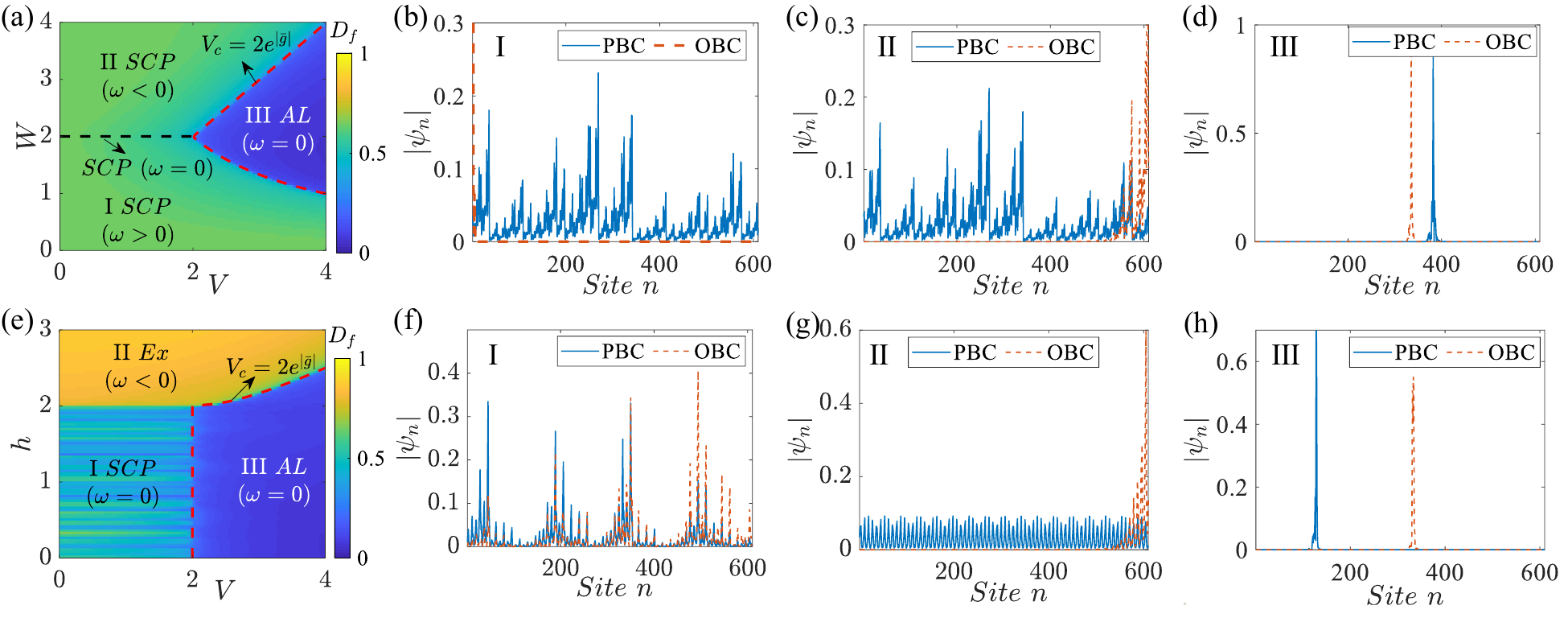}
     \caption{(a) Phase diagram in the $V$–$W$ plane for $h=0$ under the PBC. Three phases are identified: SCP with opposite winding numbers (I, II) and an AL phase (III). The red dashed line marks the analytical localization transition boundary and the black dashed line marks the zero-winding skin critical state. (b–d) The distribution of eigenstate for the three phases in (a) under PBC and OBC. (e) Phase diagram in the $V$–$h$ plane at $W=2$ under PBC, showing a zero‑winding SCP (I), an extended phase with negative winding number (II), and an AL phase (III). The red dashed line marks the analytical localization transition boundary. (f–h) Corresponding eigenstate distribution for the phases in (e) under PBC and OBC.}
    \label{figS4}
\end{figure}

\subsection{E. Effects of on-site quasiperiodic potential on the 1D skin critical phase}

In this section, we study the effects of an on-site quasiperiodic potential on the 1D skin critical phase (SCP) and show its robustness against the quasiperiodic potential. By introducing the on-site quasiperiodic potential, the 1D Hamiltonian is rewritten as
\begin{equation}
     \hat{H}_{1D}'= J\sum_{n=1}^{L-1} \left( e^{-g_n} \hat{c}_{n}^\dagger \hat{c}_{n+1} + e^{g_n} \hat{c}_{n+1}^\dagger \hat{c}_{n} \right)+ V\cos(2\pi\alpha n) + \hat{H}_B,
\label{eq34}
\end{equation}
where $g_n = \ln(|W \cos(2 \pi \beta n+\varphi)+h \cos( \pi n)|)$ and $V$ is the strength of quasiperiodic potential. When the quasiperiodic potential is sufficiently strong, the system undergoes an Anderson localization transition. For the quasiperiodic model with nonreciprocal hopping, we can derive that the inverse localization length is given by
\begin{equation}
    \Lambda^{-1} = \log\!\left(\frac{V}{2e^{|\bar{g}|}}\right),
    \label{eq35}
\end{equation}
where $\bar{g}$ is the mean value of $g_n$ in the large $L$ limit. At the localization transition point, the localization length $\Lambda$ diverges, i.e., $\Lambda^{-1}=0$. This condition yields the critical point at
\begin{equation}
    W=W_c = 2e^{|\bar{g}|}.
    \label{eq36}
\end{equation}
This gives rise to the exact boundary for the localization transition between the SCP and localized phase in the presence of quasiperiodic potential and nonreciprocal hopping.

We first fix $h=0$ and plot the phase diagram on the $V$-$W$ parameter plane under the PBC, as shown in Fig. \ref{figS4}(a). The phase diagram exhibits three distinct phases. Phases I and II are both characterized by a fractal dimension $D_f \approx 0.64$, corresponding to the SCP with different winding numbers: the winding numbers $\omega > 0$ in Phase I and $\omega < 0$ in Phase II. States on the boundary between the two phases remain skin critical states but possess a zero winding number $\omega=0$. Phase III displays a vanishing fractal dimension and belongs to the Anderson-localized (AL) phase. The eigenstates of each phase under both the PBC and OBC are illustrated in Figs. \ref{figS4}(b-d). Eigenstates in Phases I and II are bulk-interface skin critical states under the PBC, while become left-skin states and right-skin states under the OBC, respectively. Eigenstates in Phase III are fully localized under both the PBC and OBC.

We further fix $W = 2$ and plot the phase diagram on the $V$-$h$ parameter plane under the PBC and obtain a phase regime with the SCP of zero winding number, as shown in Fig. \ref{figS4}(e). This diagram also exhibits three phases: the SCP phase with $\omega = 0$ (phase I), the extended phase with $D_f \to 1$ and $\omega < 0$ (phase II), and the AL phase (phase III). The eigenstates of these phases under the PBC and OBC are presented in Figs. \ref{figS4}(f-h). In phase I with the winding number $\omega = 0$, the system is globally reciprocal yet locally non-reciprocal, and the eigenstates preserve the characteristics of the bulk-interface SCP under both boundary conditions. The eigenstates in the extended phase are extended under the PBC but become right-skin modes under the OBC, and are always localized in the AL phase under both boundary conditions.

\subsection{F. Skin critical states in open quantum systems}

In this section, we demonstrate that the skin critical states can also emerge in open quantum systems. The coupling between the system and the environment introduces dissipation and quantum jumps. The dynamics of such an open quantum system is generally governed by the Lindblad master equation
\begin{equation}
    \frac{d}{dt} \hat{\rho} = -i[\hat{H}, \hat{\rho}] + \sum_{n} \left( \hat{L}_{n} \hat{\rho} \hat{L}_{n}^{\dagger} - \frac{1}{2} \{ \hat{L}_{n}^{\dagger} \hat{L}_{n}, \hat{\rho} \} \right),
    \label{eq37}
\end{equation}
where $\hat{\rho}$ is the density operator, $\hat{H}$ describes the coherent system dynamics, and $\hat{L}_{n}$ are quantum jump operators arising from the system-environment coupling. Equivalently, the master equation can be rewritten as
\begin{equation}
    \frac{d}{dt} \hat{\rho} = -i(\hat{H}_{\text{eff}} \hat{\rho} - \hat{\rho} \hat{H}_{\text{eff}}^{\dagger}) + \sum_{n} \hat{L}_{n} \hat{\rho} \hat{L}_{n}^{\dagger}.
    \label{eq38}
\end{equation}
When the effect of the quantum jump can be disregarded, the evolution of the system is governed by an effective non-Hermitian Hamiltonian
\begin{equation}
    \hat{H}_{\text{eff}} = \hat{H} - \frac{i}{2} \sum_{n} \hat{L}_{n}^{\dagger} \hat{L}_{n}.
    \label{eq39}
\end{equation}
For the 1D lattice considered here, we choose $\hat{H}=J \sum_{n}(\hat{c}_{n}^{\dagger}\hat{c}_{n+1}+\hat{c}_{n+1}^{\dagger} \hat{c}_{n})$ and take the non-local jump operators $\hat{L}_{n}=\sqrt{\gamma_{n}}(\hat{c}_{n} + i \hat{c}_{n+1})$, which describe dissipative system–environment interactions with non-uniform couplings $\gamma_{n}$. Evaluating the dissipation term $\hat{L}_{n}^{\dagger} \hat{L}_{n} = \gamma_{n}[\hat{c}_{n}^{\dagger} \hat{c}_{n} + \hat{c}_{n+1}^{\dagger} \hat{c}_{n+1} + i(\hat{c}_{n}^{\dagger} \hat{c}_{n+1} - \hat{c}_{n+1}^{\dagger} \hat{c}_{n})]$ and substituting it into Eq. (\ref{eq39}) yields
\begin{equation}
    \hat{H}_{\text{eff}} = \sum_{n}\left[ \left( J+\frac{\gamma_{n}}{2} \right) \hat{c}_{n}^{\dagger} \hat{c}_{n+1} + \left( J-\frac{\gamma_{n}}{2} \right) \hat{c}_{n+1}^{\dagger} \hat{c}_{n} \right] - \frac{i}{2} \sum_{n} \gamma_{n} \left( \hat{c}_{n}^{\dagger} \hat{c}_{n} + \hat{c}_{n+1}^{\dagger} \hat{c}_{n+1} \right).
    \label{eq40}
\end{equation}
To obtain an effective Hamiltonian similar to the Hatano-Nelson model with quasi-periodic imaginary gauge fields in Eq. (\ref{eq24}), we set $J \pm \gamma_{n} / 2 = J e^{\pm g_{n}}$, which yields $\gamma_{n} = 2 J \tanh g_{n}$. By substituting this relation into Eq. (\ref{eq40}), we obtain the effective Hamiltonian
\begin{equation}
    \hat{H}_{\text{eff}} = \sum_{n} J \text{sech} g_{n} \left( e^{g_{n}} \hat{c}_{n}^{\dagger} \hat{c}_{n+1} + e^{-g_{n}} \hat{c}_{n+1}^{\dagger} \hat{c}_{n} \right) - i  \sum_{n} J\tanh g_{n} \left( \hat{c}_{n}^{\dagger} \hat{c}_{n} + \hat{c}_{n+1}^{\dagger} \hat{c}_{n+1} \right),
    \label{eq41}
\end{equation}
where both the non-reciprocal hopping and the on-site dissipation are modulated by the quasiperiodic imaginary gauge phase $g_n$.

\begin{figure}[htbp]
    \centering
     \includegraphics[width=0.9\linewidth]{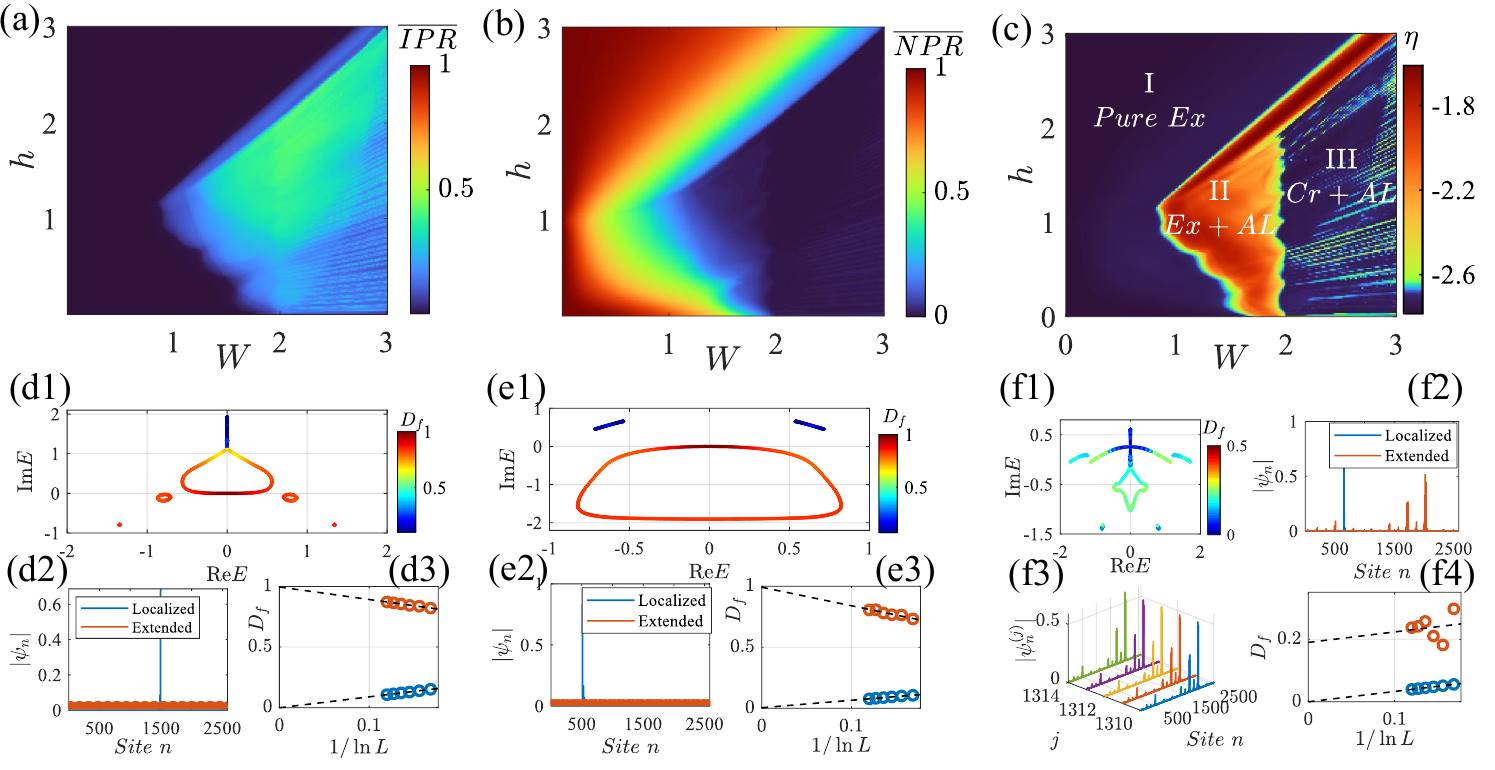}
\caption{
    (a-c) Phase diagrams on the $W$-$h$ parameter plane for  $\overline{\mathrm{IPR}}$ (a), $\overline{\mathrm{NPR}}$ (b), and $\eta$ (c) under the PBC. Region I is the pure extended phase, Regions II and III are mixed phases. (d1-d3) Analysis for Region II at $(W, h) = (1, 1)$. (d1) Complex energy spectrum colored by $D_f$. (d2) Spatial profiles of a representative extended state and a localized state. (d3) Finite-size scaling of the fractal dimension $D_f$.
    (e1-e3) Consistent analysis for a different point at $(W, h) = (2.5, 2.7)$ within Region II. (f1-f4) Analysis for Region III at $(W, h) = (2.5, 1)$. (f1) Complex energy spectrum colored by $D_f$; (f2) Spatial profiles of a critical state and a localized state. (f3) Spatial profiles of five skin critical states. (f4) Finite-size scaling of the fractal dimensions $D_f$.}
    \label{figS5}
\end{figure}

In the presence of quasiperiodically modulated on-site dissipation, the system exhibits richer localization phenomena. In addition to the emergence of skin critical states, the system also hosts mixed phases separated by mobility rings, including the coexistence of extended and localized states, as well as the coexistence of critical and localized states. To characterize the localization properties of these phases, we introduce several quantities for analysis. The inverse participation ratio (IPR) and the normalized participation ratio (NPR) for the $j$-th eigenstate are defined as
\begin{equation}
\text{IPR}^{(j)}=\sum_{n=1}^{L}\left|\psi_{n}^{(j)}\right|^{4}, \quad \text{NPR}^{(j)}=\left[L \times \text{IPR}^{(j)}\right]^{-1}.
\label{eq42}
\end{equation}
To describe the localization properties of the entire spectrum, we further average over all eigenstates to obtain the spectral-averaged IPR and NPR:
\begin{equation}
\overline{\text{IPR}}=\frac{1}{L} \sum_{j=1}^{L} \text{IPR}^{(j)}, \quad \overline{\text{NPR}}=\frac{1}{L} \sum_{j=1}^{L} \text{NPR}^{(j)}.
\label{eq43}
\end{equation}
To clearly distinguish mixed phases from pure phases, we further use a composite quantity
\begin{equation}
\eta = \log_{10}\bigl(\overline{\text{IPR}} \times \overline{\text{NPR}}\bigr).
\label{eq44}
\end{equation}

Under the PBC, we plot the localization phase diagrams of these three quantities on the $W$-$h$ parameter plane, as displayed in Figs. \ref{figS5}(a-c), respectively. According to the value of $\eta$, the parameter plane can be divided into three regions, as shown in Fig. \ref{figS5}(c). In Region I, $\eta \approx 0$, while $\overline{\text{IPR}} \approx 0$ and $\overline{\text{NPR}} \approx 1$, corresponding to a pure extended phase. Both Region II and Region III exhibit finite $\eta$, with the overall values of $\eta$ in Region II being larger than those in Region III. The results indicate that both Regions II and III are mixed phases but with distinct localization features. To determine the localization properties in these two regions, we select several representative parameter points for  detailed analysis. For Region II, we calculate the eigenvalues and eigenstates at $W = h = 1$. The resulting complex energy spectrum is shown in Fig. \ref{figS5}(d1). The fractal dimension $D_f$ of each eigenstate reveals that there is a mobility ring in the complex energy spectrum, separating extended states with non-zero winding number from localized states with zero winding number. The spatial profiles of an extended state and a localized state are displayed in Fig. \ref{figS5}(d2). The former is delocalized across the whole lattice, whereas the latter is localized. As shown in Fig. \ref{figS5}(d3), the finite-size scaling shows that their fractal dimensions approach $D_f = 1$ and $D_f = 0$ in the thermodynamic limit, respectively, confirming their localization characters. Moreover, for a strip-shaped subregion inside Region II, we consider another representative point at $W=2.5$ and $h=2.7$ with consistent features, as shown in Figs. \ref{figS5}(e1-e3). Therefore, Region II is identified as a mixed phase in which extended and localized states coexist. For Region III, we choose the representative parameters $W=2.5$ and $h=1$, with the complex energy spectrum shown in Fig. \ref{figS5}(f1), which exhibits a mobility ring separating critical states with non-zero winding number from localized states with zero winding number. The distributions of typical wavefunctions are shown in Fig. \ref{figS5}(f2), where the wavefunction of a critical state exhibits the self-similar fractal structure. In Fig. \ref{figS5}(f3), we display five eigenstates belonging to the non-zero winding part of the spectrum. Their spatial profiles are essentially identical. All of them are skinned to some particular interfaces while simultaneously showing multifractal features, identified as the skin critical states. Finite-size analysis in Fig. \ref{figS5}(f4) yields the fractal dimensions $D_f \approx 0.2$ and $D_f\approx0$ for the skin critical state and the localized state in the thermodynamic limit, which confirms that Region III is a mixed phase of skin critical and localized states.

\end{document}